**ARTICLE**

# Cartilage-binding antibodies induce pain through immune complex–mediated activation of neurons

Alex Bersellini Farinotti[1]*, Gustaf Wigerblad[1]*, Diana Nascimento[1]*, Duygu B. Bas[1], Carlos Morado Urbina[1], Kutty Selva Nandakumar[2,3], Katalin Sandor[1], Bingze Xu[2], Sally Abdelmoaty[1], Matthew A. Hunt[1], Kristina Ängeby Möller[1], Azar Baharpoor[1], Jon Sinclair[1], Kent Jardemark[1], Johanna T. Lanner[1], Ia Khmaladze[2], Lars E. Borm[4], Lu Zhang[5], Fredrik Wermeling[6], Mark S. Cragg[7], Johan Lengqvist[6], Anne-Julie Chabot-Doré[8], Luda Diatchenko[8], Inna Belfer[9], Mattias Collin[10], Kim Kultima[11], Birgitta Heyman[5], Juan Miguel Jimenez-Andrade[12], Simone Codeluppi[4], Rikard Holmdahl[2,3]**, and Camilla I. Svensson[1]**

**Rheumatoid arthritis–associated joint pain is frequently observed independent of disease activity, suggesting unidentified pain mechanisms. We demonstrate that antibodies binding to cartilage, specific for collagen type II (CII) or cartilage oligomeric matrix protein (COMP), elicit mechanical hypersensitivity in mice, uncoupled from visual, histological and molecular indications of inflammation. Cartilage antibody–induced pain-like behavior does not depend on complement activation or joint inflammation, but instead on tissue antigen recognition and local immune complex (IC) formation. smFISH and IHC suggest that neuronal *Fcgr1* and *Fcgr2b* mRNA are transported to peripheral ends of primary afferents. CII-ICs directly activate cultured WT but not FcRγ chain–deficient DRG neurons. In line with this observation, CII-IC does not induce mechanical hypersensitivity in FcRγ chain–deficient mice. Furthermore, injection of CII antibodies does not generate pain-like behavior in FcRγ chain–deficient mice or mice lacking activating FcγRs in neurons. In summary, this study defines functional coupling between autoantibodies and pain transmission that may facilitate the development of new disease-relevant pain therapeutics.**

## Introduction

The molecular dialog between the immune system and nociceptive neurons is a fundamental aspect of both acute and chronic pain. In particular, the contribution of the adaptive immune system has recently come into focus. Reports show that autoantibodies against specific neuronal proteins increase the excitability of nociceptors without involvement of other inflammatory factors (Klein et al., 2012; Dawes et al., 2018). For instance, autoantibodies against components of the voltage-gated potassium channel complex isolated from patients with Morvan's syndrome can directly elicit hyperexcitability in specific subsets of nociceptive neurons and cause neuropathic pain (Klein et al., 2012; Dawes et al., 2018). Similarly, autoantibodies have been suggested to cause pain in rheumatoid arthritis (RA). Recent studies demonstrate that individuals can be seropositive for RA-associated autoantibodies such as rheumatoid factor and anti-citrullinated protein antibodies for several years before clinical onset of the disease (Rantapää-Dahlqvist et al., 2003), and antibodies present during early stages of arthritis can interact with joint cartilage and collagen type II (CII; Pereira et al., 1985; Haag et al., 2014). During the period immediately before diagnosis, individuals frequently suffer from joint pain, often without signs of joint inflammation (de Hair et al., 2014). Furthermore, pain still persists in a sizable proportion of RA patients for whom other RA symptoms, including joint inflammation, are medically controlled (Taylor et al., 2010). Thus, joint pain uncoupled from apparent disease activity is a pervasive problem and represents a fundamental gap in our mechanistic understanding of pain in autoimmune disorders.

[1]Department of Physiology and Pharmacology, Karolinska Institutet, Stockholm, Sweden; [2]Section for Medical Inflammation Research, Department of Medical Biochemistry and Biophysics, Karolinska Institutet, Stockholm, Sweden; [3]School of Pharmaceutical Sciences, Southern Medical University, Guangzhou, China; [4]Department of Medical Biochemistry and Biophysics, Karolinska Institutet, Stockholm, Sweden; [5]Department of Medical Biochemistry and Microbiology, Uppsala University, Uppsala, Sweden; [6]Department of Medicine, Karolinska Institutet and Karolinska University Hospital, Stockholm, Sweden; [7]Centre for Cancer Immunology, Faculty of Medicine, University of Southampton, Southampton General Hospital, Southampton, UK; [8]Alan Edwards Centre for Research on Pain, McGill University, Montréal, Quebec, Canada; [9]Office of Research on Women's Health, National Institutes of Health, Bethesda, MD; [10]Division of Infection Medicine, Department of Clinical Sciences, Lund University, Lund, Sweden; [11]Department of Medical Science, Uppsala University, Uppsala, Sweden; [12]Department of Unidad Academica Multidisciplinaria Reynosa Aztlan, Universidad Autonoma de Tamaulipas, Reynosa, Tamaulipas, Mexico.

*A. Bersellini Farinotti, G. Wigerblad, and D. Nascimento contributed equally to this paper; **R. Holmdahl and C.I. Svensson contributed equally to this paper; Correspondence to Camilla I. Svensson: camilla.svensson@ki.se; Rikard Holmdahl: rikard.holmdahl@ki.se.

A subgroup of RA patients display elevated levels of circulating and intrasynovial anti-CII antibodies around the time of RA diagnosis, though their precise frequency is debated (Clague and Moore, 1984; Pereira et al., 1985). CII is a structural protein mainly found in articular cartilage, and rodents and primates immunized with CII develop an autoimmune response and joint pathology similar to human RA (Lindh et al., 2014). The transfer of monoclonal anti-CII antibodies to rodents causes a similar pathological state (Holmdahl et al., 1986; Terato et al., 1992), which is the basis for the collagen antibody–induced arthritis (CAIA) model (Nandakumar et al., 2003). When we assessed pain-like behavior in the CAIA model, we found that mechanical hypersensitivity develops before any signs of joint inflammation and remains for weeks after inflammation has subsided (Bas et al., 2012; Agalave et al., 2014; Su et al., 2015). Anti-CII antibodies cause denaturation of collagen fibrils and loss of chondrocytes in vitro (Amirahmadi et al., 2005) and early loss of proteoglycans in vivo, without the influence of inflammation (Nandakumar et al., 2008). However, as cartilage is not innervated, the anti-CII antibodies must act on other targets to mediate pronociceptive effects in the preinflammatory stage. Thus, the aim of this study was to investigate the pronociceptive properties of anti-CII antibodies.

## Results
### Induction of pain-like behavior by anti-CII antibodies is not associated with inflammation
CAIA was induced by injection of an anti-CII mAb cocktail followed by LPS 5 d later. Cell infiltration, bone erosion, and cartilage destruction were readily detectable by day 15. We observed not only that mice displayed a reduction in tactile thresholds during the disease phase, but also that mechanical hypersensitivity was already present before visible joint inflammation, on days 3 and 5 (Fig. 1, A–C). Although no ankle-joint pathology was observed before day 5, synovitis was present in two of eight mice, with coincident arthritis scores of 5 and 13 on a scale of 1–60 (Fig. 1, D–G). No correlation was found between Von Frey pain-like behavior and arthritis scores at day 5 ($r = 0.159$, $P = 0.634$, $n = 19$).

To determine if anti-CII mAbs, in the absence of LPS, induce a low-grade inflammation capable of activating sensory neurons, joints were processed for molecular analysis of factors associated with arthritis pathology and pain signaling. While mRNA levels of tumor necrosis factor (*Tnf*), interleukins *Il-1b* and *Il-6*, prostaglandin-producing enzyme cyclooxygenase-2 (*Cox2*), mast cell proteases (*Mcpt4*), and matrix metalloproteases (MMPs) *Mmp2*, *Mmp9*, and *Mmp13* were significantly increased at day 15 of the CAIA model, none of them were elevated 5 d after injection of the anti-CII mAb cocktail compared with saline controls (Fig. 1, H and I). Furthermore, changes in MMP activity were examined using MMPsense. An increase in fluorescent signal was detected in the paws 15 d after the injection of anti-CII mAb cocktail, but again no differences were observed between antibody-injected mice and saline controls on day 5, suggesting that MMPs were not activated at this time point (Fig. 1, J and K). Taken together, these results suggest that factors other than innate inflammatory and extracellular matrix remodeling mediators drive anti-CII mAb–induced mechanical hypersensitivity before onset of joint inflammation. Thus, in the subsequent studies, we focused on the early phase of the CAIA model (days 0–5; before LPS injection) in order to explore the mechanisms by which anti-CII mAbs induce pain-like behavior before inflammation.

### Pain-like behavior is apparent as early as 2 d after injection of anti-CII antibodies
When assessed daily, we found that the anti-CII mAb cocktail induced a significant reduction in tactile thresholds by day 2 compared with saline-injected animals (Fig. 2 A). None of the mice displayed signs of joint inflammation before day 4, and only 3 of 14 developed mild signs of joint inflammation by day 5, characterized by arthritis scores ranging from 5 to 13 (Fig. 2, B and C). Reduction in locomotor activity has been used as a surrogate of pain-related mobility impairment in rodents (Cho et al., 2013). When assessed during the third night (12-h period) after antibody injection, a reduction in total movement and rearing activity was detected (Fig. 2 D), suggesting that anti-CII mAbs decrease voluntary and spontaneous locomotion in mice before signs of inflammation. In contrast, when we performed the inverted grid test, none of the mice injected with anti-CII mAbs (day 5) or saline ($n = 10$/group) fell from the fully turned grid, indicating that both groups displayed similar grip and muscular strength.

### Antibody epitope recognition, but not pathogenicity, is important in early anti-CII antibody–induced pain-like behavior
To investigate whether the arthritogenic potency of different anti-CII mAbs correlates with their pronociceptive potency, we injected the antibodies individually and measured mechanical sensitivity. All four antibodies, but not the isotype control antibodies, induced similar degrees of mechanical hypersensitivity when injected individually (Fig. 2, E–G), as well as in combination (Fig. 2 A). Injection of the anti-CII mAb M2139 alone, the most arthritogenic antibody in the cocktail, also reduced total movement compared with the isotype IgG2b control antibody, although the difference in rearing between the groups did not reach statistical significance ($P = 0.054$; Fig. 2 H). M2139 induced mechanical hypersensitivity at different doses (Fig. 2 I), which lasted up to 21 d following a single injection, even with doses of M2139 that failed to induce joint inflammation at any time point (Fig. 2, J and K).

### Complement factor C5 and changes in cartilage structure do not contribute to early pain-like behavior
To examine if anti-CII mAbs induce nociception through activation of the complement cascade, PMX53, a cyclic peptide C5aR antagonist, was injected daily starting 1 d before administration of the anti-CII mAb cocktail (day 0). The C5aR antagonist failed to reverse antibody-induced changes in mechanical hypersensitivity or locomotor activity (Fig. 3, A–C). Furthermore, the degree of mechanical hypersensitivity and reduction in locomotion were not different between B10Q.C5* mice lacking functional C5 and WT mice after injection of anti-CII mAbs (Fig. 3, D–F). As previously shown, the CIIF4 antibody binds to CII but does not lead to cartilage damage (Burkhardt et al., 2002; Nandakumar et al., 2008; Croxford et al., 2010). Injection of CIIF4 antibody also induced robust mechanical hypersensitivity comparable to the other anti-CII mAbs tested (Fig. 3 G). These experiments indicate that the mechanism responsible for anti-CII antibody–mediated nociception is independent of C5 (and thereby terminal/lytic complement) or changes in cartilage structure.

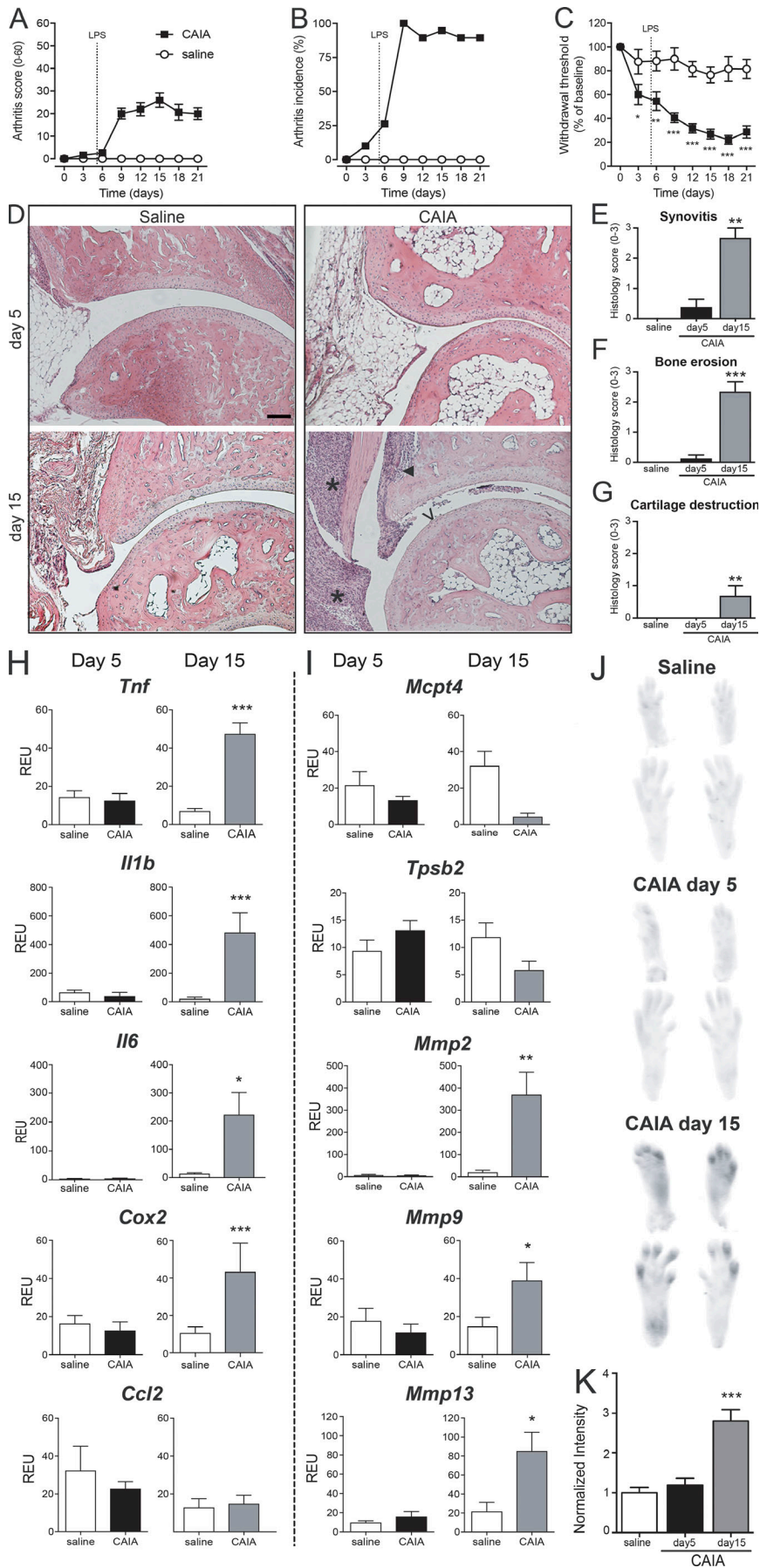

Figure 1. **Injection of anti-CII antibodies induces pain-like behavior before visual, histological, and molecular signs of arthritis. (A–C)** B10.RIII mice injected with anti-CII mAbs (n = 19; saline controls n = 17) started developing joint inflammation around day 6 (A). On day 9, all animals displayed signs of arthritis (B). Mechanical hypersensitivity (C) was observed already on days 3 and 5, before onset of arthritis, and persisted throughout day 21. **(D)** Representative H&E histology of B10.RIII mouse ankle joints collected 5 and 15 d after injection of anti-CII mAbs. While an inflammatory infiltrate, bone erosion, and cartilage serration were visible on day 15, no signs of joint pathology was detectable on day 5 or in saline controls. Scale bar represents 100 µm. *, ▼, and V point to signs of synovitis, bone erosion, and cartilage destruction, respectively. **(E–G)** Scores for inflammatory hallmarks as synovitis (E), bone erosion (F), and loss of cartilage (G) revealed mild ankle joint pathology in two of eight mice day 5 and prominent signs in all mice day 15 (n = 5). Control mice represent pooled time-matched saline-injected mice (n = 4+4). **(H and I)** Quantitative PCR analysis of joint extracts showed a significant increase in mRNA levels of most of the inflammatory factors investigated at day 15 (n = 7), while none of them were elevated at day 5 of CAIA (n = 6), compared with saline controls (n = 5; H and I). **(J and K)** Activation of MMPs was significantly increased only after 15 d of CAIA, while no changes were detected at day 5 (n = 3/group, B10.RIII mice). Data are presented as mean ± SEM. *, P < 0.05; **, P < 0.01; ***, P < 0.001 compared with saline controls. REU, relative expression unit.

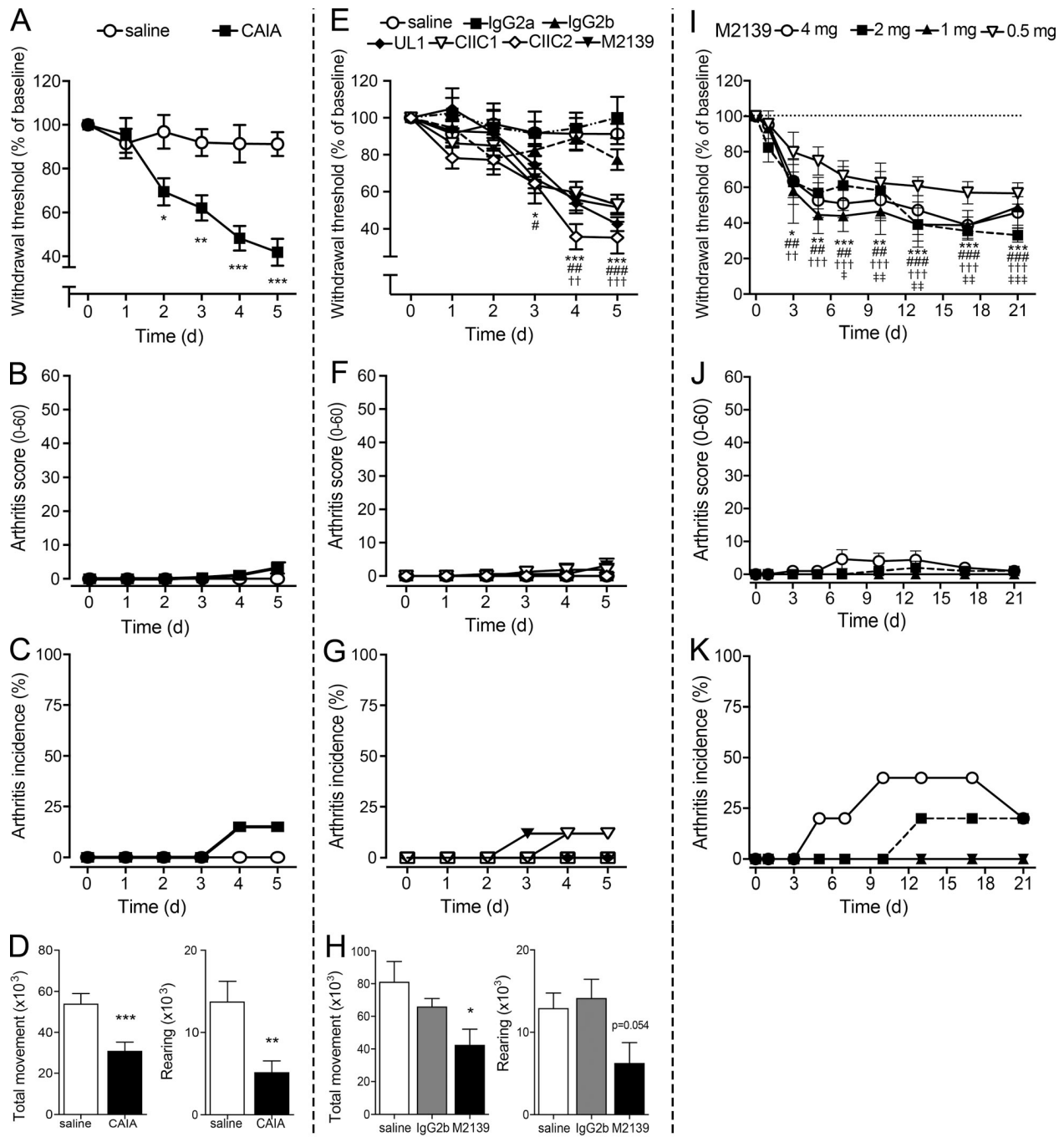

Figure 2. **Anti-CII antibodies injected either as a cocktail or as individual antibodies induce mechanical hypersensitivity and reduce locomotion before inflammation. (A–C)** Anti-CII mAbs (n = 10) induced mechanical hypersensitivity as early as 2 d after injection (A) compared with saline controls (n = 9) in B10.RIII mice. Arthritis scores (B) and incidence (C) were not detectable until day 4 and remained very low also on day 5. **(D)** Total movement (left) and rearing (right) significantly decreased in B10.RIII mice injected with the anti-CII mAb cocktail (n = 15), compared with controls (n = 19). **(E–G)** When injected individually, the four mAbs (M2139, UL1, CIIC1, and CIIC2) induced mechanical hypersensitivity similarly to the cocktail (E; n = 5–9/group, B10.RIII mice). No considerable signs of inflammation (F and G) were detected. **(H)** Total movement (left) and rearing (right) were reduced in M2139 mAb–injected B10.RIII mice (n = 5), compared with saline (n = 5) or isotype control (n = 5). **(I–K)** Injection of M2139 mAb–induced mechanical hypersensitivity for 21 d (I) even at doses that did not induce visual signs of inflammation (J and K; n = 5, B10.RIII mice). Axes in Fig. 2 (A and E) are interrupted to make the difference between groups clearer to visualize. Data are presented as mean ± SEM. *, P < 0.05; **, P < 0.01; ***, P < 0.001 compared with saline controls. For Fig. 2 E, significance is shown with symbols: * for M2139 and CIIC2 mAbs, # for CIIC1, and † for UL1 compared with saline controls. For Fig. 2 I, significance is shown with symbols: * for 4 mg M2139, # for 2 mg M2139, † for 1 mg M2139, and ‡ for 0.5 mg M2139 compared with each respective baseline.

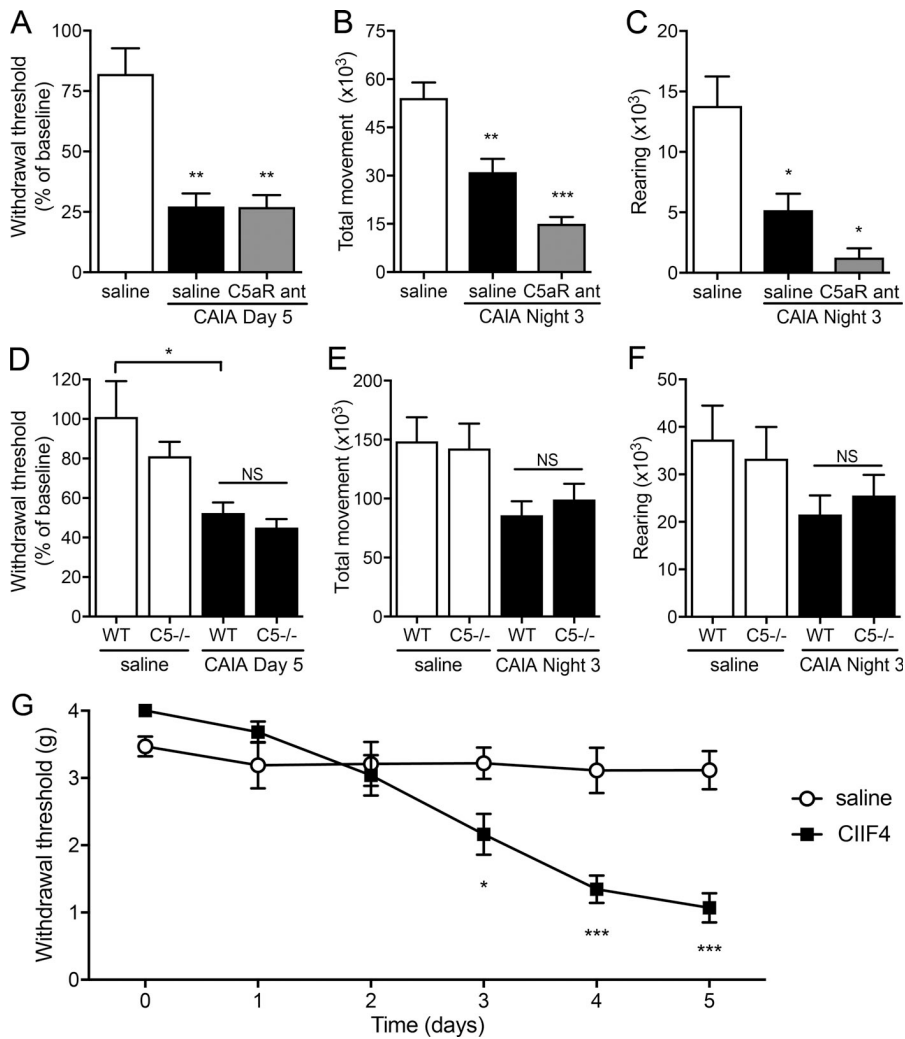

Figure 3. **Anti-CII antibody–induced pain-like behavior is not mediated by complement activation or cartilage destruction. (A)** Injection of the C5a-receptor antagonist PMX53 (C5aR ant; n = 5) did not prevent anti-CII mAb–induced mechanical hypersensitivity (n = 4, B10.RIII mice) compared with vehicle (saline)-injected controls (n = 7, B10.RIII mice). **(B and C)** Antagonizing the C5a-receptor (n = 5, B10.RIII mice) did not prevent anti-CII mAb–induced reduction in total movement (B) and rearing (C) compared with saline controls (n = 19, B10.RIII mice). **(D–F)** Complement 5–deficient ($C5^{-/-}$) mice developed mechanical hypersensitivity (D; n = 5) and displayed a reduction in total movement (E) and rearing (F; n = 4) comparable to WT B10Q mice (n = 6–8) after injection of anti-CII mAbs. **(G)** B10.RIII mice injected with the nonarthritogenic CIIF4 antibody (n = 8) developed mechanical hypersensitivity from day 3 after injection compared with saline controls (n = 7). Data are presented as mean ± SEM. *, P < 0.05; **, P < 0.01; ***, P < 0.001 compared with saline controls.

**FcγRs are present in mouse sensory neurons**

As an alternative pronociceptive mechanism, we explored interactions between CII immune complexes (CII-ICs) and neurons. We examined expression of FcγRs in mouse sensory neurons using several different techniques. First, we observed the mRNA expression of all four *Fcgrs* (*Fcgr1*, *Fcgr2b*, *Fcgr3*, and *Fcgr4*) in mouse dorsal root ganglion (DRG) via gene expression microarrays (Fig. 4 A), which were subsequently confirmed by quantitative real-time PCR (Fig. 4 B). These data are in line with the publicly available resource DRG XTome database (Ted Price laboratory, University of Texas at Dallas, Dallas, TX), which shows the presence of *Fcgr* mRNA in mouse DRGs (Fig. 4 C; Ray et al., 2018). Using single-molecule fluorescence in situ hybridization (smFISH), we detected mRNA molecules for *Fcgr1*, *Fcgr2b*, and *Fcgr3* in both neuronal (colocalizing with NeuN) and non-neuronal cells, but failed to detect *Fcgr4* (Fig. 4 D). Quantification of single mRNA molecules for each receptor in individual sensory neurons plotted by area of neuronal soma showed the highest expression of *Fcgr1* in neurons (Figs. 4 D and S1). Using Western blotting for protein analysis, Fcγ receptor (FcγR) I was detected in DRG and spleen (positive control) homogenates from WT but not from FcRγ-chain$^{-/-}$ mice, which lack cell surface expression and signaling of all activating FcγRs (I, III, and IV; Fig. 4 E; Takai et al., 1994; Nimmerjahn et al., 2005). As FcγRIIb and FcγRIII antibodies did not work for Western blotting, we examined their presence in full DRGs lysates with high-performance nano–liquid chromatography/tandem mass spectrometry (nanoLC-MS/MS) proteomics, revealing the presence of FcγRIIb based on the identification of two unique peptides originating from FcγRIIb (Fig. 4 F), along with two peptides that are shared between FcγRIIb and FcγRIII (data not shown).

Cellular localization was examined via immunohistochemistry (IHC). FcγRI immunoreactivity was detected in DRGs of WT BALB/c and C57BL/6 mice (Figs. 5 A and S2), colocalizing with Iba1-positive resident macrophages (Fig. 5 B), but not satellite cells or neurons (lack of colocalization with vimentin and tropomyosin receptor kinase A [TrkA], respectively; Fig. S2). As FcγRI is expressed in the soma of rat DRG neurons (Qu et al., 2011, 2012; Jiang et al., 2017), we verified our finding by using anti-FcγRI antibodies from several vendors (Figs. S2 and S3) and using DRG sections from FcRγ-chain$^{-/-}$ mice as a negative control. Three of four antibodies tested on DRG sections labeled resident macrophages in WT mice with no detectable signal in FcRγ-chain−/− mice, and the fourth displayed nonspecific labeling in all sections (Figs. 5 A and S3).

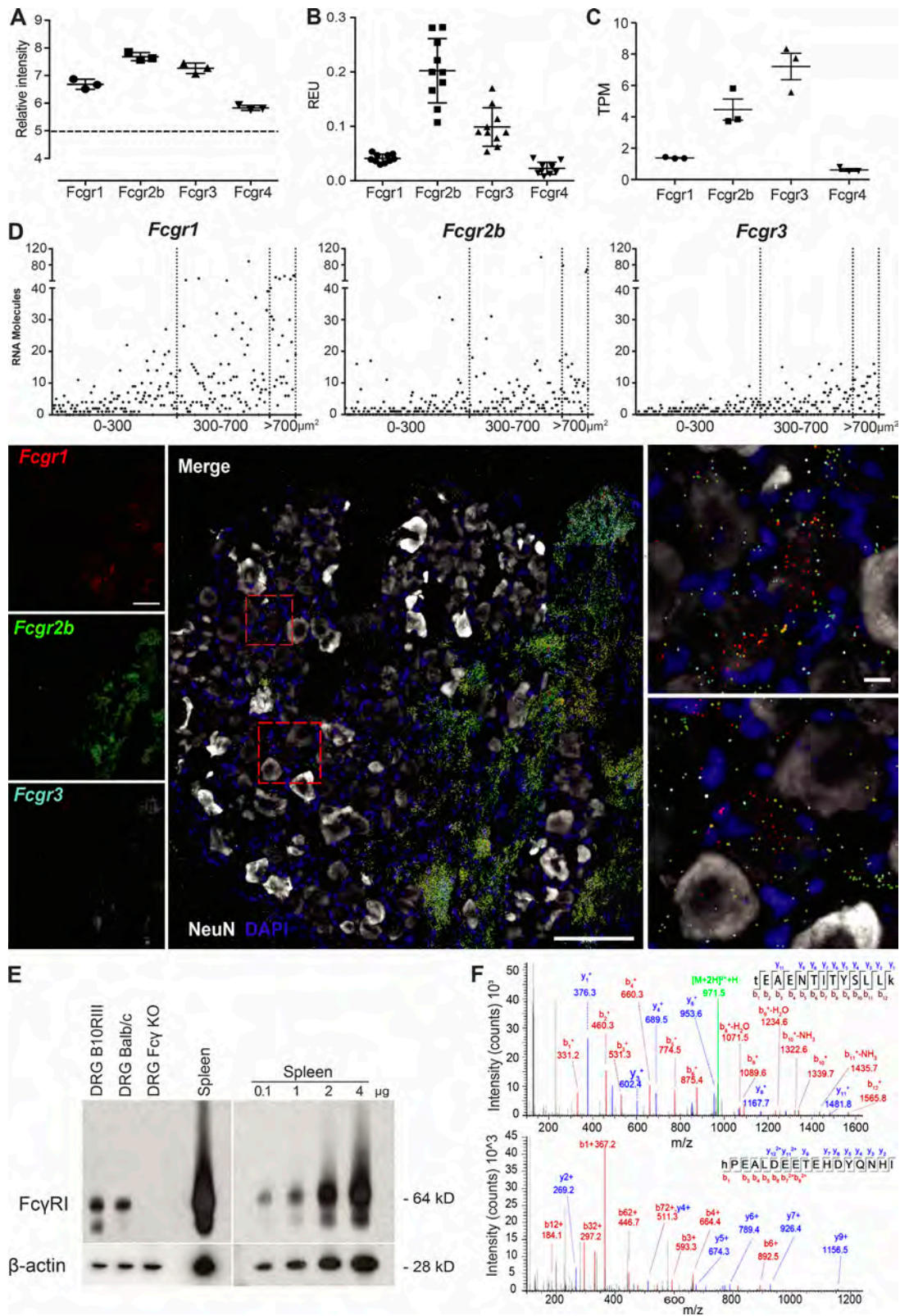

Figure 4. **FcγRs are expressed in mouse DRG neurons. (A)** Microarray data showed mRNA for *Fcgr1–4* in CBA mouse DRG (n = 3). **(B)** Quantitative PCR showed mRNA for *Fcgr1–4* in B10.RIII mouse DRG (n = 10). **(C)** Publicly available RNA sequencing of C57BL/6 mouse DRGs show the presence for *Fcgr1–4* (n = 3). **(D)** smFISH showed mRNA molecules for *Fcgr1*, *Fcgr2b*, and *Fcgr3* in BALB/c mouse DRG colocalizing with NeuN. Scatter graph shows number of mRNA molecules in neuronal soma. Scale bars represent 100 μm and 10 μm in close-up images. **(E)** FcγRI protein expression was detected by Western blotting in DRGs from WT BALB/c and B10.RIII mice, but not from FcRγ-chain$^{-/-}$ mice. **(F)** Proteomic analysis identified peptides specific for FcγRIIb in BALB/c mouse DRG. See also Figs. S1 and S4.

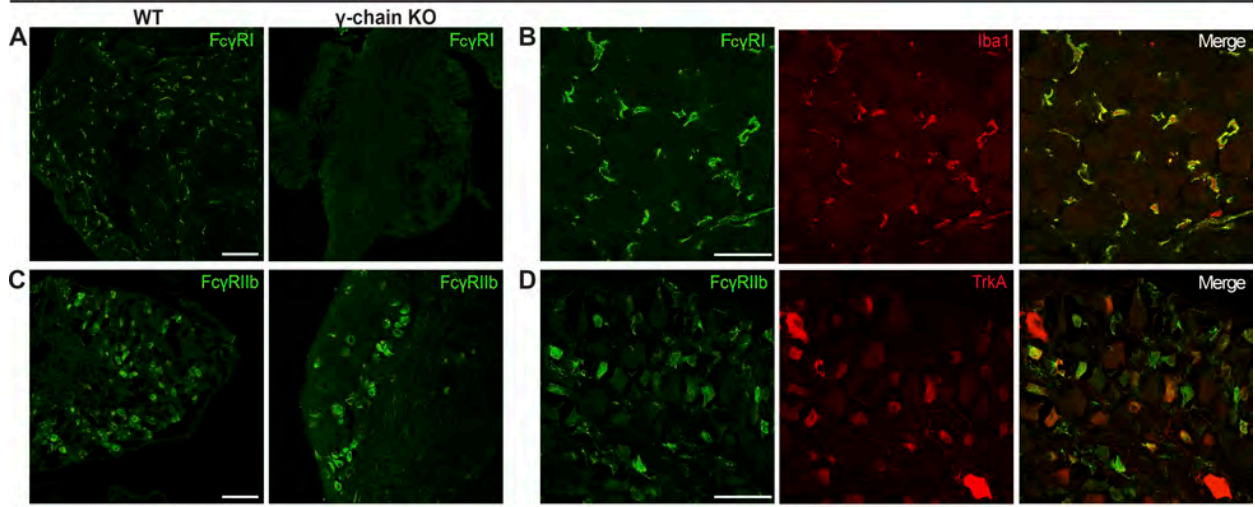
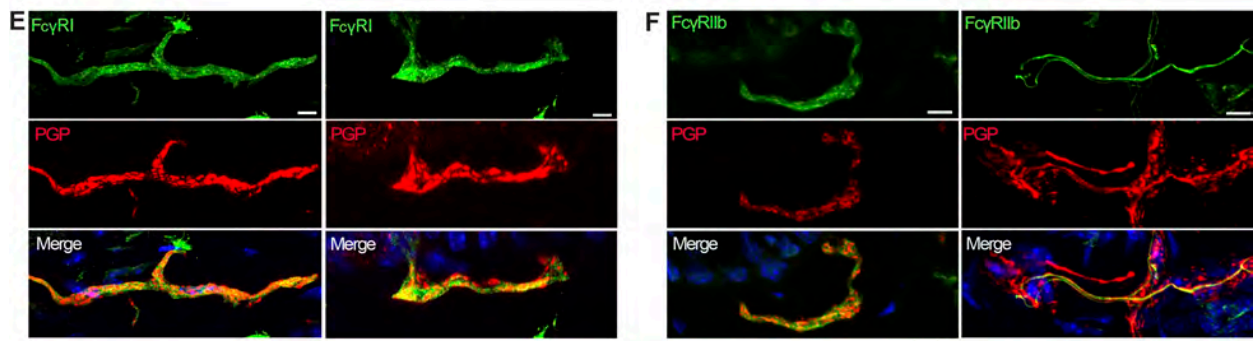
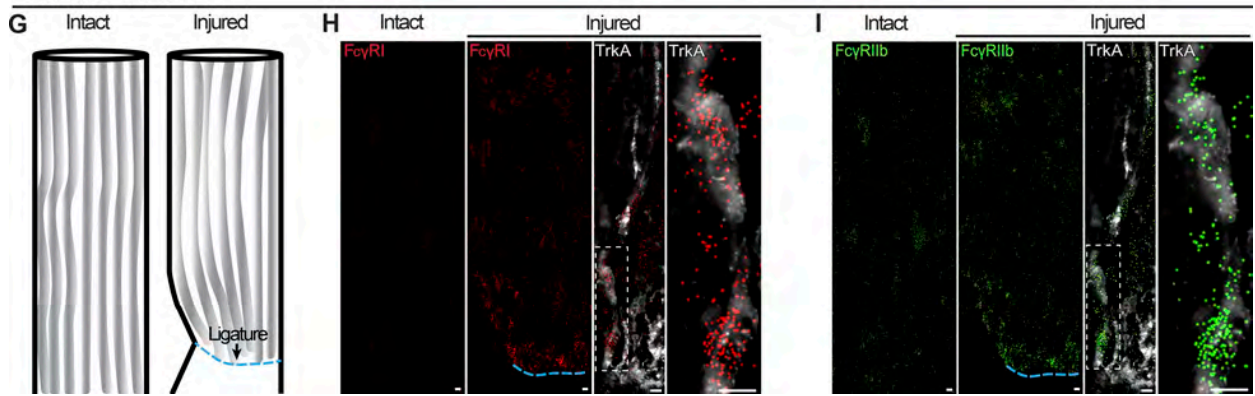

Figure 5. **FcγRI and FcγRIIb are expressed in the DRG and in nerve fibers in the skin. (A and B)** FcγRI immunoreactivity was detected in WT BALB/c DRGs, but not in FcRγ-chain$^{-/-}$ mice (A), colocalizing with Iba1-positive resident macrophages (B). Scale bars represent 100 µm and 50 µm, respectively. **(C and D)** FcγRIIb immunoreactivity was detected in BALB/c mouse DRG and retained in FcRγ-chain$^{-/-}$ mice (C), colocalizing with TrkA-positive neurons (D). Scale bars represent 100 µm and 50 µm, respectively. **(E and F)** FcγRI (E) and FcγRIIb (F) immunoreactivity was detected in PGP9.5-positive nerve fibers in BALB/c mouse glabrous skin. Scale bars represent 5 µm. **(G–I)** smFISH on BALB/c mouse sciatic nerves after ligation (G) revealed accumulation of mRNA molecules for *Fcgr1* (H) and *Fcgr2b* (I) proximal to the site of ligation (ipsilateral), while barely any signal was found in the contralateral intact nerve. Scale bars represent 10 µm. See also Figs. S2, S3, and S4.

In contrast, FcγRIIb protein expression was detected in the soma of DRG neurons, confirmed by colocalization with neuronal marker TrkA (Fig. 5, C and D). As expected, FcγRIIb signal was still present in FcRγ-chain−/− mice (Fig. 5 C). No immunoreactivity was detected for FcγRIII and FcγRIV (Fig. S2). Both FcγRI and FcγRIIb immunoreactivity were detected in glabrous skin sections, and although FcγRI was expressed exclusively in macrophages in the DRG, both FcγRI and FcγRIIb colocalized with neuronal marker PGP9.5 in the skin (Fig. 5, E and F). Both receptors were also detected in nonneuronal cells in the skin.

Due to the lack of FcγRI protein expression in the cell bodies of DRG neurons, the presence of FcγRI immunoreactivity in the end structures of PGP9.5-positive neurons and the high levels of Fcgr1 mRNA in nonneuronal, nonnuclear areas in the DRG that contain mainly fiber tracts (Fig. S4), we examined whether Fcgr1 mRNA may be transported down the axon for local translation. smFISH performed after ligation of the sciatic nerve (Fig. 5 G) showed that Fcgr1 and Fcgr2b mRNA molecules accumulated proximal to the ligature within fiber tracts, compared with the contralateral nerve, indicating axonally transported mRNA (Fig. 5, H and I). In conclusion, both FcγRI and FcγRIIb are expressed in sensory neurons; however, while FcγRIIb protein was detected both in the DRG cell body and axons, FcγRI protein was detected only in the peripheral axon in vivo.

### CII-IC activates cultured DRG neurons, causing increased intracellular [Ca$^{2+}$] and inward current

Primary DRG neuronal cell cultures were used for in vitro experiments. FcγRI and FcγRIIb immunoreactivity colocalized with the neuronal marker (βIII-tubulin); FcγRIIb expression was more pronounced in the cell bodies, and FcγRI immunoreactivity was more prominent in the axons and neurites (Fig. 6 A). Immunoreactivity for FcγRIII and FcγRIV was not detected (data not shown). Primary DRG neurons stimulated with CII-IC exhibited an increased intracellular [Ca$^{2+}$] signal in 247 cells (22.1%) of 1,119 viable neurons (KCl responding), while stimulation with monomeric control IgG2b evoked response in <1% of viable neuronal cells (Fig. 6 B). The average percent change in signal intensity after CII-IC was 99.6 ± 0.1% (mean ± SEM).

We then performed electrophysiological recordings on a subpopulation of nociceptive neurons that express transient receptor potential vanilloid 1 (TRPV1) receptors. Capsaicin (0.5 µM), a potent TRPV1 agonist, was added at the end of each experiment to verify neuronal population and viability. In total, 114 cells were patched, and ionic currents were recorded in whole-cell voltage clamp mode. Of the 114 cells, 52 cells showed an inward current in response to capsaicin (48%). Stimulation of capsaicin-responding neurons with CII-IC evoked inward currents in 42% (22 of 52, 19% of total cells), while stimulation with IgG2b failed to evoke inward currents in any of the 18 neurons assayed (Fig. 6 C). The average amplitude of responses after CII-IC application was 49.38 ± 8.89 pA (mean ± SEM), and the majority of the cells displayed currents >25 pA (data not shown). Moreover, we performed a similar experiment applying a generic IC (mouse anti-rat IgGs and, as antigen, rat IgGs) and found that 10 of 46 total cells (22%) patched showed inward currents in response to IgG-IC.

### CII-IC induces calcitonin gene–related peptide (CGRP) release in primary DRG cultures from WT but not FcRγ-chain−/− mice

CGRP is a neuropeptide, expressed in nociceptive neurons, released upon various noxious stimuli (van Rossum et al., 1997). A bell-shaped dose–response relationship was observed upon DRG stimulation with 0.1–10 µg/ml of CII-IC, with the largest CGRP release induced by 1 µg/ml of CII-IC (Fig. 6 D). In contrast, CGRP levels in the DRG culture supernatants were not elevated in response to stimulation with monomeric anti-CII mAbs, CII, or control IgG2b antibody (Fig. 6 D). Furthermore, while primary DRG neuronal cultures established from FcRγ-chain−/− mice responded to stimulation with the positive control, capsaicin, stimulation with CII-IC did not induce CGRP release (Fig. 6 E). Finally, in Ca$^{2+}$ imaging experiments, there was no difference in the percentage of neurons responding to CII-IC in cultures established from FcγRIII−/− (10 of 85 viable neurons) and WT (8 of 92 viable neurons) mice (8.7% and 11.8%, respectively). These results suggest that the neuronal high-affinity FcγRI, rather than the low-affinity FcγRIIb, is responsible for CII-IC–induced release of CGRP in vitro.

### Intra-articular (i.a.) injection of IC induces pain-like behavior

To investigate whether CII-IC induces pain-like behavior in vivo, we injected CII-IC into the i.a. space of the ankle joint. I.a. injections of CII-IC (Fig. 7 A), as well as a general IC (Fig. 7 E), elicited mechanical hypersensitivity in the ipsilateral paw 1 or 3 h after injection. Furthermore, i.v. injection of an antibody to cartilage oligomeric matrix protein (COMP; Geng et al., 2012), a major noncollagenous component of cartilage, and i.a. injection of COMP-IC induced pain-like behavior in the absence of inflammation (Fig. 7, F–H). In contrast, neither i.a. nor i.v. injection of CII-IC or anti-CII mAbs induced mechanical hypersensitivity in FcRγ chain−/− mice (Fig. 7, B–D). Finally, systemic injection of anti-CII mAbs induced mechanical hypersensitivity in FcγRIV−/− mice, which persisted at 30 d despite these mice being resistant to induction of CAIA (Fig. 7, I–K). These data suggest that activating FcγRs, and FcγRI in particular, are critical for development of IC-mediated pain-like behavior in vivo.

### Intact and glycosylated CII antibodies are required for pronociceptive effect in vivo

Fab fragments of the four anti-CII mAbs did not induce signs of mechanical hypersensitivity or altered locomotion upon i.v. injection (Fig. 8, A–C), indicating the Fc portion as necessary for development of pain-like behavior. Moreover, to confirm a role of FcγR interaction, we used the endoglycosidase EndoS known to cleave the Fc glycan so that the antibodies have reduced affinity for Fc receptors and reduced capacity to form ICs (Nandakumar et al., 2007, 2018). EndoS-treated anti-CII mAb M2139 injected i.v. did not induce mechanical hypersensitivity or reduced locomotion compared with control mice, even though a robust change in behavior was observed in mice receiving intact M2139 anti-CII mAbs (Fig. 8, E–G). Similarly, EndoS-treated anti-CII mAb cocktail did not reduce withdrawal thresholds (Fig. 8 D). These observations, taken together, indicate that the binding of Fab to CII alone is not capable of activating nociceptors and that glycosylation of Fc is required.

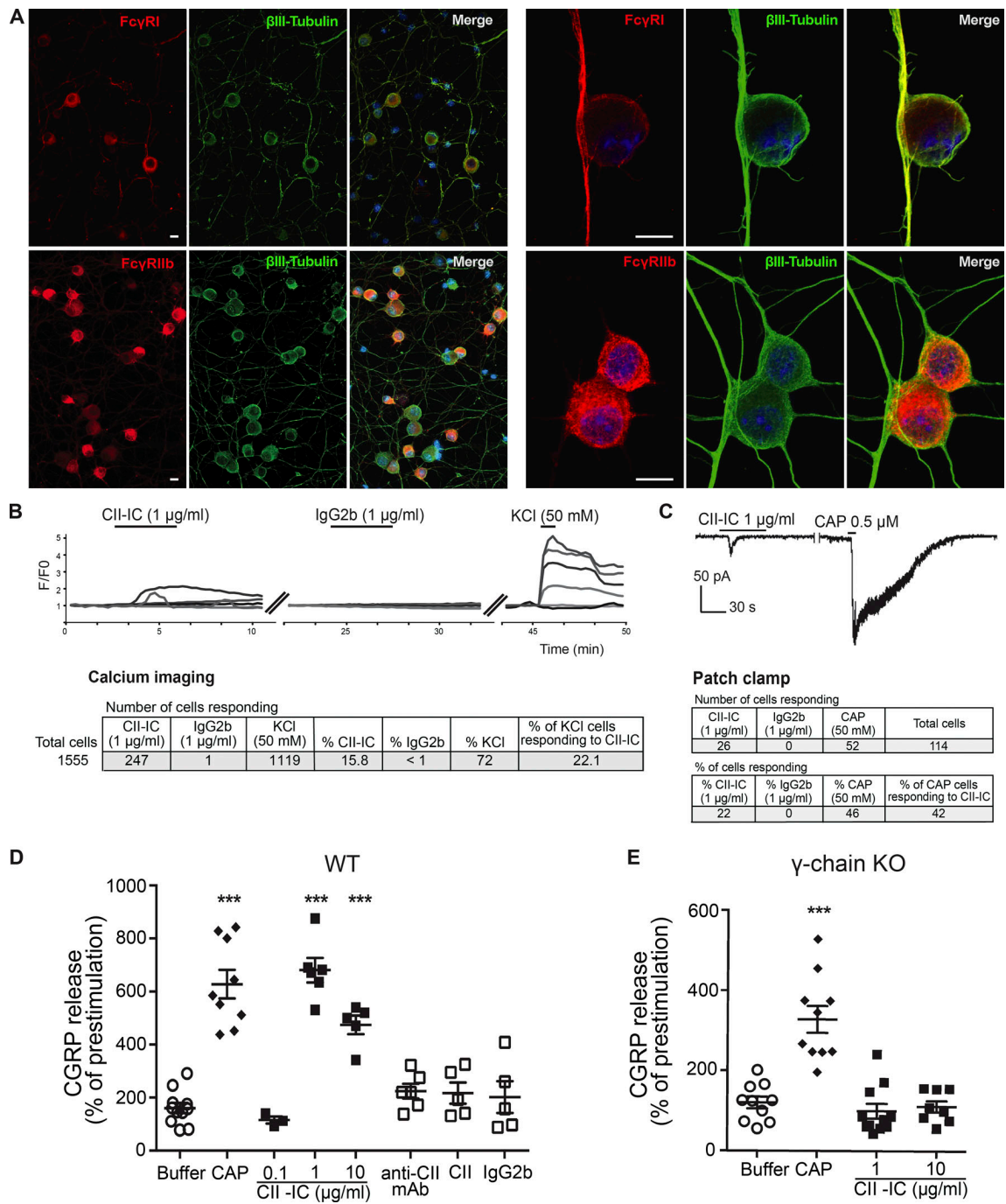

Figure 6. **CII-IC stimulation of DRG cell cultures leads to increased neuronal excitability. (A)** FcγRI and FcγRIIb are expressed in BALB/c mouse DRG neurons in culture as shown by colocalization with βIII-tubulin. Scale bars represent 10 µm. **(B and C)** CII-IC stimulation of BALB/c mouse DRG cell cultures resulted in increased intracellular [$Ca^{2+}$] signal (B) and also evoked positive inward currents (C). **(D and E)** CII-IC stimulation evoked CGRP release in DRG cell cultures from WT BALB/c mice (D) but not from FcRγ-chain$^{-/-}$ mice (E). Capsaicin (CAP) was used as positive control. Data are presented as mean ± SEM. ***, $P < 0.001$.

**FcγRs on neurons and not hematopoietic cells are responsible for anti-CII antibody–induced pain-like behavior**

We used chimeric mice to investigate the contribution of FcγRI to anti-CII antibody–induced pain in hematopoietic cells compared with nonhematopoietic cells (including neurons). Mice were irradiated to deplete hematopoietic cells and then transplanted with bone marrow (BM) from either WT or FcRγ-chain$^{-/-}$ mice. Irradiated WT mice that received WT BM were used as a control. Three groups of mice ($_{WT}$-KO, $_{KO}$-WT, and $_{WT}$-WT) were injected with either saline or anti-CII mAbs and monitored for mechanical thresholds for 6 d. Mice expressing activating FcγRs solely on hematopoietic cells and not neurons or other nonhematopoietic cells ($_{WT}$-KO) were protected from anti-CII antibody–induced mechanical hypersensitivity, while mice lacking FcγRs on myeloid cells ($_{KO}$-WT) developed mechanical hypersensitivity indistinguishable from control mice ($_{WT}$-WT; Fig. 8, H–J). While these results do not exclusively test the role of activating FcγRs on neurons, they indeed support such a link, as FcγRs on immune cells are not critical for induction of anti-CII mAb–mediated pain-like behavior.

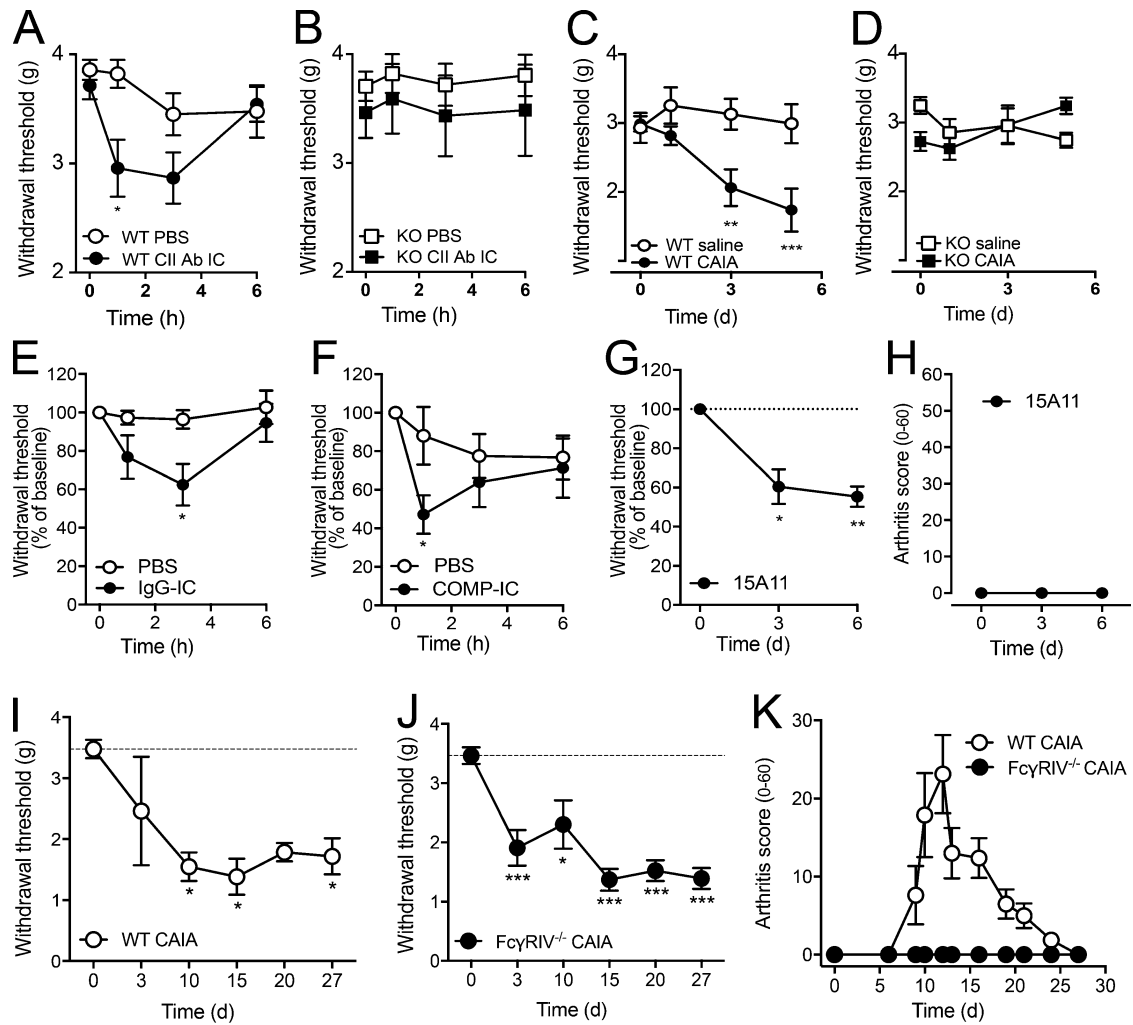

Figure 7. **Different ICs promote pain-like behavior in vivo, and FcγRIV$^{-/-}$ mice develop mechanical hypersensitivity despite lack of CAIA.** **(A and B)** I.a. injection of CII-IC–induced mechanical hypersensitivity in WT BALB/c mice ($n$ = 14–21/group; A) but not in FcRγ-chain$^{-/-}$ mice ($n$ = 8–10/group; B). **(C and D)** Systemic administration of anti-CII mAbs evoked mechanical hypersensitivity in WT BALB/c mice ($n$ = 8/group; C) but not FcRγ-chain$^{-/-}$ mice ($n$ = 8/group; D). **(E)** I.a. injection of IgG-IC induced mechanical hypersensitivity in WT BALB/c mice ($n$ = 6/group). **(F)** I.a. injection of COMP-IC induced mechanical hypersensitivity in WT C57BL/6 mice ($n$ = 7/group). **(G and H)** Systemic administration of anti-COMP mAb evoked mechanical hypersensitivity in WT BALB/c mice ($n$ = 5; G), in the absence of any signs of inflammation (H). **(I–K)** Systemic injection of anti-CII mAbs induced pain-like behavior in both WT C57BL/6 mice ($n$ = 4; I) and FcγRIV$^{-/-}$ mice ($n$ = 6; J), even if no signs of inflammation were observed in the latter (K). Axes in Fig. 7 (A–D) are interrupted to make the difference between groups clearer to visualize. Data are presented as mean ± SEM. *, $P < 0.05$; **, $P < 0.01$; ***, $P < 0.001$ compared with saline/PBS controls.

**Activating FcγRs are expressed by human sensory neurons**
Data from the publicly available DRG XTome database (Ted Price laboratory, University of Texas at Dallas, Dallas, TX) show the presence of *Fcgr* mRNA in human DRGs, with *Fcgr3A* being the most highly expressed (Fig. 9 A; Ray et al., 2018). Using IHC, we examined protein expression of activating FcγRs in human DRGs ($n$ = 4) and found that FcγRI expression did not colocalize with the neuronal marker NeuN but was present in cells with a macrophage-like morphology (Fig. 9, B and C). Using antibodies against FcγRIIa, FcγRIIIb, and FcγRIIIa/b, we were able to detect a positive signal only from FcγRIIIa/b, which colocalized with NeuN-positive neurons as well as nonneuronal macrophage-like cells in the human DRGs (Fig. 9, D and E).

## Discussion

In the present study, we have explored the mechanisms by which anti-CII antibodies induce pain-like behavior before induction of arthritis. We provide evidence that these antibodies trigger pain-like behavior before any visual, histological, or molecular signs of inflammation, independently of complement factor C5 and changes in cartilage integrity. By using modified anti-CII antibodies and transgenic or chimeric mice, we established that, in addition to epitope recognition, interaction with neuronal FcγRI is critical for the pronociceptive properties of anti-CII antibodies. Finally, the presence of FcγRIII in human DRG neurons suggests the translation potential of this work. While further studies are warranted, our findings support a role for neuronal FcγRs in autoimmune pain conditions.

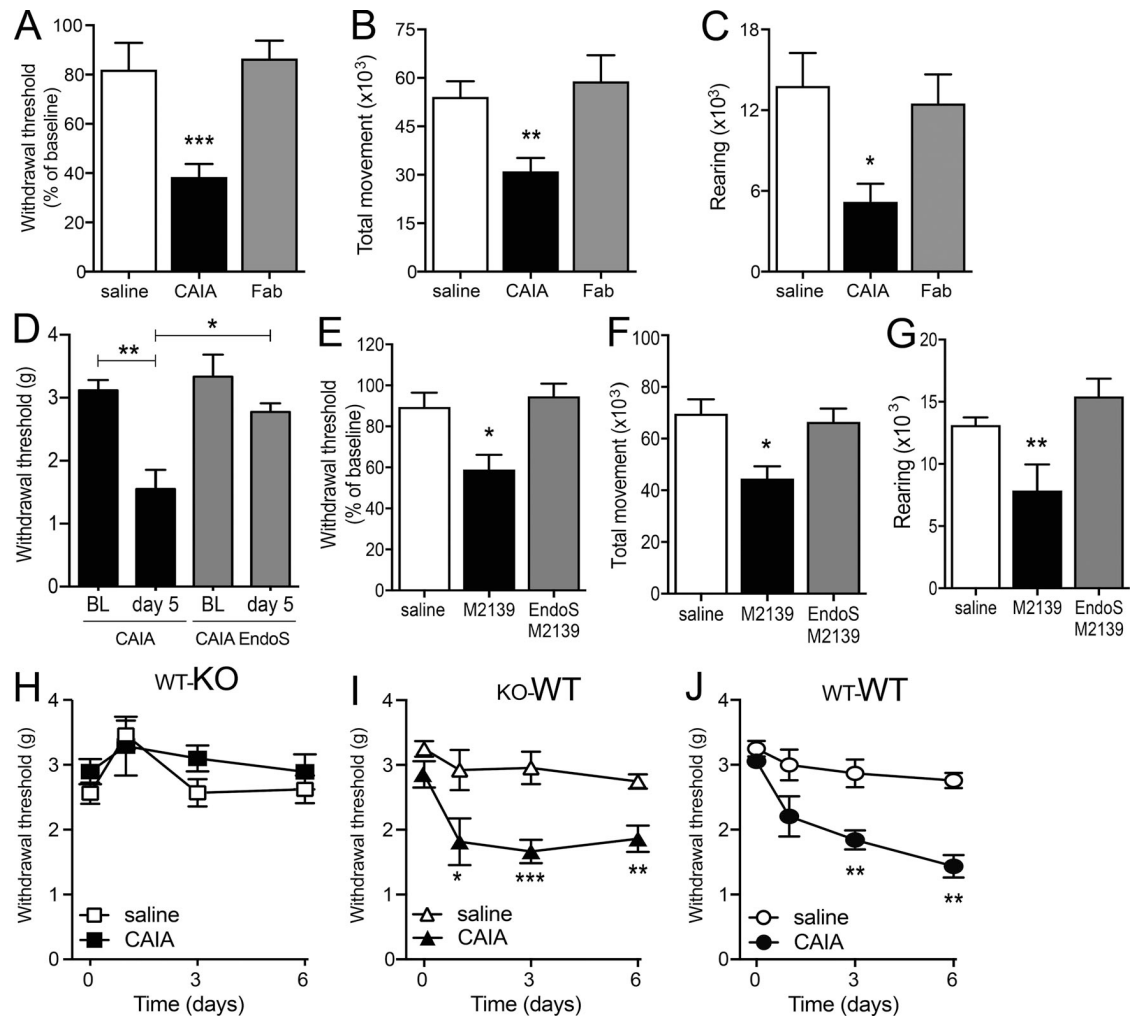

Figure 8. **The pronociceptive properties of anti-CII antibodies are dependent on the Fc region, glycosylation, and interaction with FcγRI in the joint.** **(A–C)** B10.RIII mice injected with anti-CII mAb Fab fragments ($n = 8$) did not develop mechanical hypersensitivity (day 5; A) compared with CAIA ($n = 13$) and control ($n = 7$) mice. They also did not show reduction in total movement (B) or rearing (C; night 3, $n = 8–19$/group). **(D)** B10.RIII mice injected with EndoS-treated anti-CII mAbs did not develop mechanical hypersensitivity ($n = 3$/group). BL, baseline. **(E–G)** B10.RIII mice injected with EndoS-treated anti-CII mAb M2139 did not develop mechanical hypersensitivity (day 5; E) or display a reduction in locomotor activity (F and G; night 3, $n = 6–7$/group). **(H–J)** BALB/c mice lacking activating FcγRs in myeloid cells ($_{KO}$-WT; I) developed mechanical hypersensitivity after injection of anti-CII mAbs compared with controls ($_{WT}$-WT; J). In contrast, mice lacking activating FcγRs in nonmyeloid cells ($_{WT}$-KO), including neurons, were protected (H; $n = 8–9$/group). Data are presented as mean ± SEM. *, $P < 0.05$; **, $P < 0.01$; ***, $P < 0.001$ compared with controls.

CII antibodies readily bind joint cartilage in vivo, forming ICs, which are likely crucial for the attraction of inflammatory cells and represent a key step in arthritis development. Thus, we initially hypothesized that the early pain-like behavior following injection of anti-CII antibodies is mediated by local soluble CII-IC, inducing a low-grade, local inflammation, not detectable as swelling. However, we did not find any indications of coupling between inflammatory processes and nociception during the first 5 d after antibody injection in the joints. Even though activation of inflammatory cells is often indicated as necessary for induction of RA pathogenesis, arthritogenic antibodies can directly cause cartilage destabilization both in vitro and in vivo, preceding the onset of arthritis, independently of inflammation (Nandakumar et al., 2008). We asked if associated actions could also drive pronociceptive processes. To test this, we used CIIF4, an antibody that, unlike the other anti-CII mAbs used in this study, binds CII but lacks arthritogenicity. In fact, this antibody is protective when given together with other anti-CII antibodies both in vitro and in vivo (Nandakumar et al., 2008; Croxford et al., 2010). Remarkably, CIIF4 induced pronounced mechanical hypersensitivity. Thus, the processes associated with anti-CII antibody–induced cartilage loss unlikely mediate pain-like behavior induced by the antibodies. Together, these experiments led us to conclude that neither inflammation, terminal/lytic complement, nor cartilage breakdown are mechanistic explanations for the pain-like behavior observed in the early phase of CAIA. Prompted to explore alternative mechanisms, we turned our attention to direct actions of anti-CII antibodies on peripheral neurons.

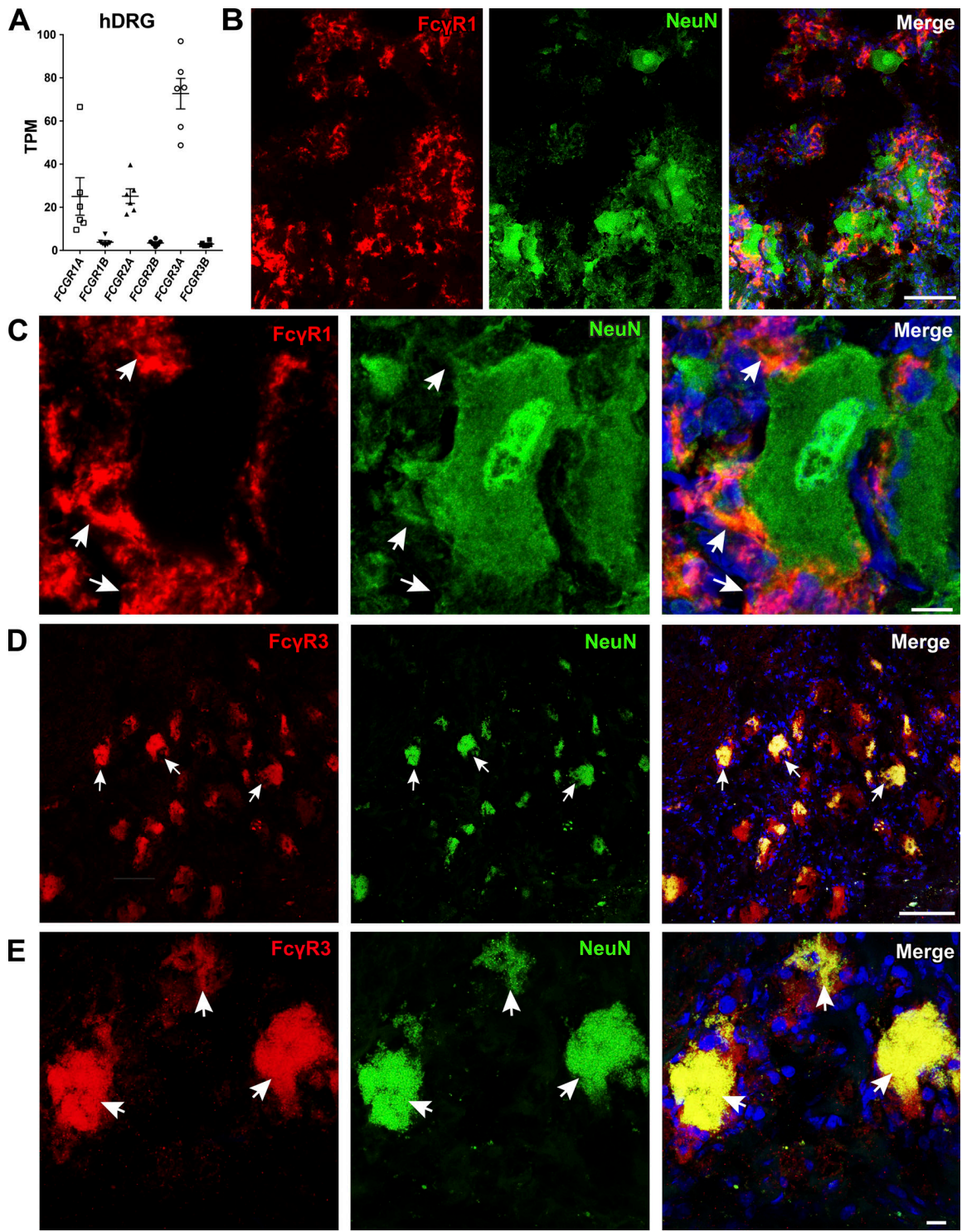

Figure 9. **FcγRI and FcγRIII are expressed in human DRG. (A)** Publicly available data show the presence of *Fcgr* mRNA in human DRGs (n = 6). *Fcgr3a* is the most highly expressed. **(B and C)** FcγRI immunoreactivity was detected in human DRGs (n = 4). The lack of colocalization with NeuN and the morphology of the FcγRI-positive cells suggest FcγRI expression in resident macrophages, similarly to mice (scale bars represent 100 µm and 10 µm in close-up images). White arrows indicate FcγRI-positive cells, which are negative for NeuN. **(D and E)** Immunoreactivity for the activating FcγRIII in human DRGs (n = 4) colocalized with the neuronal marker NeuN (scale bars represent 100 µm and 10 µm, respectively). White arrows point to double-positive neurons (FcγRIII and NeuN).

To examine if anti-CII antibodies have a direct action on nociceptors, we added the antibodies in monomeric form to DRG

neurons in culture. While the cells responded to positive controls, no increase in inward currents, intracellular [Ca$^{2+}$], or release of the pain-associated neuropeptide CGRP were observed in the presence of anti-CII or control antibodies. This is not surprising, as CII displays a very narrow tissue distribution. Except for hyaline cartilage in joints, it is found in the ear, larynx, trachea, vitreous of the eye (Eyre, 1991), and thymus (Raposo et al., 2018). We also did not detect any anti-CII positivity via Western blots from DRG lysates or primary DRG cultures (data not shown). Hence, we concluded that there are no CII epitopes present on the neuronal membrane and that a direct neuronal action of monomeric anti-CII antibodies via the Fab region is unlikely. However, the Fc region of IgG antibodies in the context of ICs can activate FcγRs. Intriguingly, several studies have shown expression of FcγRI in rat sensory neurons (Andoh and Kuraishi, 2004; Qu et al., 2011, 2012), and here we report that in addition to FcγRI, mouse DRG neurons express FcγRIIb in vitro and in vivo. Anti-CII antibodies in IC with CII evoked inward currents and increased intracellular [Ca$^{2+}$] along with release of the pain-associated neuropeptide CGRP in mouse-cultured primary DRG neurons. Thus, the presence of immune cells was not necessary for IC-mediated activation of nociceptive neurons, which supports the notion of a direct link between antibodies and regulation of neuronal excitability. Our results are consistent with previous findings showing that IgG-ICs increase intracellular Ca$^{2+}$ levels, membrane depolarization, and release of substance P from cultured DRG neurons (Andoh and Kuraishi, 2004; Qu et al., 2011; Jiang et al., 2017). Neuronally expressed FcγRI has been coupled to the activation of the cation channel TRPC3 through a signaling pathway involving Syk, phospholipase C, and the inositol triphosphate receptor (Qu et al., 2012). In previous studies examining the effect of IC activation of neuronal FcγRs, exogenous antigens (normal mouse IgG or OVA) have been used to generate ICs with rat anti-mouse IgG or anti-OVA IgG, respectively. In the current study, we employed a model that mimics the early phase of RA by relying on IC formation between anti-CII antibodies and soluble CII fragments, both of which are present in the synovial fluid (Lohmander et al., 2003; Yoshida et al., 2006). In fact, anti-CII IgGs are thought to be locally produced in RA patients, as antibody titers are often higher in synovial fluid compared with serum (Rowley et al., 1987; Lindh et al., 2014). We have previously demonstrated that after i.v. injection in mice, anti-CII mAbs reach the joint and bind cartilage after ∼24 h (Jonsson et al., 1989). COMP is expressed predominantly in cartilage, and COMP released from cartilage is associated with development of RA (Saxne and Heinegård, 1992). Anti-COMP mAbs induced pain-like behavior very early, before any signs of inflammation in mice after i.v. injection, as well as after i.a. injection in IC formation. Thus, we speculate that systemic injection of monomeric antibodies that bind antigens in the joint, e.g., soluble CII fragments or COMP, leads to local formation and accumulation of IC, which activates FcγRI on sensory neurons at concentrations that are lower than those required for the induction of inflammation. In line with this hypothesis, when we injected preformed IC i.a., the mechanical hypersensitivity developed faster compared with i.v. injection of monomeric antibodies (hours compared with days), still in the absence of visual signs of inflammation. Furthermore, we observed neuronal responses in DRG cultures only when the antibodies were applied as preformed ICs: monomeric CII mAbs were without effect, as the antigen is not present in our in vitro system, and there is no IC formation. Thus, while the antigen is critical for IC formation, the IC-FcγR interaction on sensory neurons may represent a more general pain mechanism.

We have previously shown that human IgG is not detectable in the spinal cord 7 d after i.v. injection to naive mice (Wigerblad et al., 2016). Thus, the primary site of the pronociceptive actions of anti-CII and anti-COMP antibodies is most likely peripheral rather than central. However, prolonged activation of primary afferents often leads to spinal sensitization, an important component of pain chronicity. While outside the scope of the current work, future studies exploring such aspects of neuronal FcγR-mediated hypersensitivity are important.

While FcγR expression in rat DRGs has been explored, very little information is available with regard to FcγR neuronal expression and associated pronociceptive function in mice. Thus, we carefully mapped mRNA and protein expression of the FcγRs in mice and found that *Fcgr1*, *Fcgr2b*, and *Fcgr3* mRNA was readily detectable in mouse DRG, both in the soma of sensory neurons and in nonneuronal cells. In naive rats, only FcγRI is present in DRGs, and protein expression is detectable exclusively in neurons (Qu et al., 2011, 2012). In contrast, we found both FcγRI and FcγRIIb protein in naive mouse DRGs, a finding that we confirmed using different techniques. Strikingly, mouse FcγRI protein expression was not detectable in the soma of DRG neurons, but instead was seen in resident DRG macrophages. FcγRIIb, on the other hand, was present in neuronal cell bodies. While the expression pattern of FcγRI has not been previously explored in the mouse sensory nervous system, expression in motor neurons (Mohamed et al., 2002) has been suggested.

The surprisingly high number of *Fcgr1* mRNA molecules in DRG fiber tracts caused us to hypothesize it may be axonally transported for local translation. Indeed, the machinery for mRNA translation can be found along the sensory axons, where local translation has been shown to regulate peripheral nociceptor plasticity (Jiménez-Díaz et al., 2008; Price and Géranton, 2009; Obara et al., 2012). By being locally translated, proteins are believed to display more specific targeting. While further work is necessary to determine if FcγRs are locally translated in nociceptors, we did find that ligation of the sciatic nerve resulted in accumulation of *Fcgr* mRNA molecules proximal to the ligature site, and protein expression of FcγRI, as well as FcγRIIb, is present in nerve fibers in the skin. Thus, it is an intriguing possibility that changes in neuronal FcγR expression may be a specific regulatory modulation in response to injury or inflammation, allowing for an increased capacity to react to ICs by enhanced neuronal excitability.

Previous work showed that intraplantar injection of OVA-ICs induces pain-like behavior (Jiang et al., 2017). Bringing this into the context of arthritis in the joint, we showed that i.a. injection of CII-IC, COMP-IC, and IgG-IC rapidly induced mechanical hypersensitivity in WT mice, but CII-IC failed to do so in FcRγ-chain$^{-/-}$ mice. Even though the site of neuronal synthesis differs between mouse and rat, our studies indicate that in naive mice as well, FcγRI is present at peripheral neuronal terminals, enabling the sensory nervous system to respond directly to the presence of ICs.

Our experiments using FcRγ-chain$^{-/-}$ mice and modified anti-CII antibodies that retain their ability to bind CII but either lack the Fc region or have a reduced affinity for FcγRs show that the Fc–FcγR interaction is critical for development of CII antibody–induced pain-like behavior. These experiments do not rule out that pain-like behavior is the consequence of IC activation of inflammatory cells. However, the lack of FcγRI expression on satellite cells and the fact that CII-IC failed to induce CGRP release in cultures generated from FcRγ-chain$^{-/-}$ mice, but still increased intracellular [Ca$^{2+}$] signal in cultures from FcγRIII$^{-/-}$ mice, strongly suggest that CII-IC can act on FcγRI on sensory neurons and drive neuronal excitation. Furthermore, previous work showed that mast cells are not involved in this process in rats (Jiang et al., 2017), and we found that mice depleted of activating FcγR in myeloid cells still developed early CAIA hypersensitivity. Conversely, mice lacking activating FcγR on nonmyeloid cells, including neurons, were protected from developing mechanical hypersensitivity after injection of anti-CII antibodies. Thus, accumulating evidence points to a critical role of neuronally expressed FcγR in the induction of pain before inflammation.

While we found protein expression of both FcγRI and FcγRIIb in peripheral axons of nociceptors in vivo and in vitro, our experiments do not indicate that FcγRIIb is coupled to enhancement of neuronal excitability. Using FcRγ-chain$^{-/-}$ mice, which lack all activating FcγRs but still express FcγRIIb, was sufficient to reduce the pronociceptive actions of cartilage-associated antibodies injected systemically and locally (i.a. as IC) as well as to prevent IC-evoked release of CGRP in primary neuronal DRG cultures. Nevertheless, the presence of FcγRIIb in sensory neurons is interesting and warrants further investigation, as the receptor could be linked to inhibitory mechanisms. Furthermore, as FcγRIII and FcγRIV deficiency did not reduce CII-IC–induced increase in intracellular [Ca$^{2+}$] or anti-CII antibody–induced pain-like behavior, respectively, we conclude that neuronal FcγRI is the most likely receptor responsible for the pronociceptive properties of cartilage-associated antibodies before inflammation.

The expression pattern of FcγRs in human DRGs was hitherto unexplored. Interestingly, in accordance with our observations in mice, FcγRI protein expression is not localized to neuronal cell bodies but to nonneuronal cells in human DRGs, which, based on morphology and localization, appear to be resident macrophages. Further work is warranted to determine if FcγRI is transported and locally translated in humans. Of note, we found protein expression of another activating FcγR, FcγRIII, in human DRG NeuN-positive neurons. Although the FcγRIII antibody used does not differentiate between FcγRIIIA and FcγRIIIB, we did not detect any signal with antibodies specific for FcγRIIIB, and as *FCGR3A* mRNA levels are 30-fold higher than *FCGR3B* in human DRGs, FcγRIIIA seems the most likely receptor. Thus, FcγRIIIA is an interesting target for further human studies. Considerable cell type–specific expression profile differences exist between mouse, rat, and human FcγRs, which certainly calls for caution when interpreting the translational potential of these data. Nevertheless, it is an intriguing possibility that human sensory neurons also respond to ICs through similar neuronal-FcγR pathways, as FcγRs represent a novel potential therapeutic target for pain in conditions with an autoimmune component.

In summary, the present study supports a novel view on how autoantibodies can act as pronociceptive factors. Local formation of soluble ICs has the potential to serve important roles in both the induction and maintenance of pain via mechanisms mediated by direct interactions between ICs and neuronally expressed FcγRI. This study shows that CII-ICs, which are highly correlated with early RA and joint pathology, serve as key triggers for pain behavior in the early phase of the disease. Indeed, our studies point to a functional coupling between autoantibodies and pain transmission, even in the absence of inflammation, and open new avenues for decoding pain mechanisms in autoimmune diseases.

## Material and methods

### Animals

All the experiments were approved by the local ethics committee for animal experiments in Sweden (Stockholm Norra Djurförsöksetiska nämnd). For some experiments, BALB/c, C57BL/6, and CBA mice were purchased from Charles River and Janvier Laboratories. The B10.RIII, B10Q strains, B10Q.C5* transgenic mice (mice with congenic 2-bp deletion in the complement 5–encoding gene making it nonfunctional; Johansson et al., 2001), FcγRIII$^{-/-}$ (founder from the Jackson Laboratory; Hazenbos et al., 1996), and FcγRIV$^{-/-}$ (Nimmerjahn et al., 2010) mice were bred at the Karolinska Institutet. FcγRIII$^{-/-}$ mice and C5* mice were bred on B10Q background and FcγRIV$^{-/-}$ mice on C57BL/6 background for >10 backcrosses. BALB/c WT and FcRγ chain$^{-/-}$ mice (lacking the activating receptors FcγRI, III, and IV; Takai et al., 1994; Nimmerjahn et al., 2005) were backcrossed for 12 generations to Balb/c and bred at the National Veterinary Institute, Uppsala, Sweden. For experiments involving transgenic mice, we used homozygous WT littermates mice as controls, except for experiments with FcRγ chain$^{-/-}$ mice. The WT control and FcRγ chain$^{-/-}$ mice originate from the same breeding line but were maintained as homozygous mice in parallel. Both male and female mice 12–20 wk of age were used, and all mice were housed in standard cages (three to five per cage) in a climate-controlled environment maintaining a 12-h light/dark cycle with access to food and water ad libitum. This study conforms to the Animal Research: Reporting of In Vivo Experiments guidelines.

### Antibodies and drugs

The antibody cocktail used for induction of CAIA contains equal amounts of four arthritogenic anti-CII mouse mAbs: M2139 (IgG2b, J1 epitope), CIIC1 (IgG2a, C1 epitope), CIIC2 (IgG2b, D3 epitope), and UL1 (IgG2b, U1 epitope; Nandakumar and Holmdahl, 2005). The anti-CII mAb CIIF4 was used as a nonarthritogenic CII-binding antibody (Nandakumar et al., 2008;

Croxford et al., 2010). mIgG2a (mouse anti-human HLA-DRa [L243]; Abcam) and mIgG2b (mouse anti-human parathyroid epithelial cells) were used as isotype control mAbs. 15A11 was used as anti-COMP mAb (Geng et al., 2012). Antibodies were produced and purified as described earlier (Nandakumar and Holmdahl, 2005). CII-IC stock solution (1 mg/ml) was prepared by mixing anti-CII mAb cocktail (1 mg/ml) with rat CII (1 mg/ml) at a ratio of 1:1 at 37°C with gentle shaking for 30 min (Burkhardt et al., 2002). Similarly, COMP-IC and IgG-IC were prepared by mixing antibodies (15A11 anti-COMP and mouse anti-rat IgGs, respectively) with antigen (COMP and rat IgGs, respectively) at ratios of 6:1 and 1:1, respectively, at 37°C with gentle shaking for 1 h.

To hydrolyze the N-linked Fc-glycans, M2139 mAb or anti-CII mAbs cocktail were incubated with recombinant endo-β-N-acetyl-glucosaminidase (EndoS) fused to glutathione S-transferase (GST) as previously described (Collin and Olsén, 2001). Briefly, GST-EndoS in PBS was mixed with M2139 mAb or anti-CII mAbs cocktail and incubated at 37°C for 16 h. GST-EndoS was then removed using glutathione-Sepharose 4B columns (GE Healthcare). Further purification of the antibodies was done using an ion exchange column. SDS-PAGE and Lens culinaris agglutinin lectin blotting were used to confirm complete removal of GST-EndoS and efficacy of EndoS cleavage. Fab fragments were prepared from the anti-CII mAb cocktail using the Pierce Fab Preparation Kit (Thermo Fisher Scientific) according to the manufacturer's instructions.

### Experimental models and drug/antibody administration

All mAbs were injected i.v. CAIA was induced by injection of anti-CII antibody cocktail (4 mg in 150 µl saline) followed by i.p. injection of LPS (*Escherichia coli* LPS, 25 µg in 100 µl saline, 055: B5; Sigma-Aldrich) 5 d later. LPS boosts the immune activity and synchronizes the onset of disease, detectable as a rapid increase in the arthritis score and incidence of arthritis. In the experiment with FcγRIV$^{-/-}$ mice, a 5-clone CII antibody cocktail was injected i.v. (3 mg; Chondrex) followed by 50 µg LPS in 100 µl saline (O111:B4; Chondrex) on day 5, as this gives a higher arthritis incidence in C57BL/6 mice. The early phase of the CAIA model was defined as day 0–5 after injection of anti-CII antibodies (i.e., before LPS injection). CIIF4, M2139, CIIC1, CIIC2, UL1, and control IgGs were injected individually (4 mg in 150 µl saline). Also, Fab fragments and EndoS-treated antibodies corresponding to 4 mg anti-CII mAb cocktail were injected i.v in 150 µl saline. For different doses test, 0.5–4 mg of M2139 mAb was administered in 150 µl saline. The cyclic peptide C5a-receptor inhibitor (PMX53, 3 mg/kg in saline; Academia Sinica, gift from Dr. Alice Yu, University of California, San Diego, San Diego, CA) was injected s.c. 1 h before injection of anti-CII mAb cocktail and then once daily 3 h before assessment of mechanical hypersensitivity.

For i.a. injections. mice were anesthetized with isoflurane (induction, 5%; maintenance, 2.5%), and different ICs (500 ng in 5 µl PBS) were injected into the ankle joint using a 29-G needle. For BM transplantation experiments, recipient BALB/c FcRγ chain$^{-/-}$ and WT mice were irradiated with 750 rad. The following day, BM cells were harvested from donor mice by flushing the tibia and femur, and $10 \times 10^6$ cells in 0.2 ml PBS were injected i.v. in the recipient mice. Irradiated FcRγ chain$^{-/-}$ mice received WT BM cells, which generated mice with activating FcγR expression on hematopoietic cells but negative on other cells including neurons ($_{WT}$-KO). Irradiated WT mice received BM from FcRγ-chain$^{-/-}$ mice, which generated mice with activating FcγRs only on nonhematopoietic cells including neurons ($_{KO}$-WT). Irradiated WT mice receiving BM cell transfer from WT mice ($_{WT}$-WT) were used as controls. Recipient mice were injected with saline or anti-CII mAbs i.v. (4 mg in 150 µl saline) 6 wk after irradiation. Nerve ligation was established by ligating the tibial and common peroneal branches (with 6-0 silk suture) under isoflurane anesthesia. Mice received buprenorphine (0.1 mg/kg s.c.) every 12 h for 2 d after surgery.

### Assessment of arthritis

The development of arthritis in the fore and hind paws was monitored by visual inspection as described previously (Holmdahl et al., 1998; Bas et al., 2012). Briefly, visible signs of inflammation, defined as redness and swelling, were scored on a 0–60 scale by investigators blinded for the origin and treatment of the mice. Each inflamed digit was noted as 1 point, and inflammation of the metacarpus/metatarsus and ankle joint as 5 points, giving a maximum of 15 points per paw. Incidence was calculated as percentage of mice that were positive for arthritis. The degree of arthritis was also assessed by histology. Mice were deeply anesthetized with volatile isoflurane (5%) and perfused with saline followed by 4% paraformaldehyde (PFA). Hind ankle joints were postfixed in 4% PFA for 48 h, decalcified in EDTA (Sigma-Aldrich) solution for 4–5 wk, and then dehydrated in ethanol and embedded in paraffin. Sections (5 µm) were stained with H&E (Histolab) and scored by blinded investigators as previously described (Bas et al., 2012) on a scale from 0 to 3, where 0 is normal and 3 is severe synovitis, bone erosion, and/or cartilage destruction.

### Assessment of pain-like behavior

Mechanical hypersensitivity in the hind paws and reduced locomotion were used as measures of evoked and spontaneous pain-like behavior, respectively. Assessment of mechanical hypersensitivity was performed on indicated days, and locomotion was assessed during the night between days 2 and 3 (third night) after antibody injection. The investigators were blinded to the origin and treatment of the mice during behavioral assessments and data analysis.

#### Mechanical hypersensitivity

Paw withdrawal thresholds were measured using von Frey filaments. Mice were habituated to the testing cages, individual compartments on top of a wire-mesh surface (Ugo Basile), before baseline recordings. On test days, mice were first habituated to the test environment for 1 h before testing. Withdrawal thresholds were assessed by application of OptiHair filaments (Marstock OptiHair) of increasing stiffness (0.5, 1, 2, 4, 8, 16, and 32 mN, corresponding to 0.051, 0.102, 0.204, 0.408, 0.815, 1.63, and 3.26 g, respectively) to the plantar surface of the paw. A brisk withdrawal of the paw from the filament within 2–3 s was noted as a positive response. The 50% withdrawal threshold (force of the filaments necessary to produce a reaction from the animal in 50% of the applications) was calculated using the Dixon up-down method (Chaplan et al., 1994) and expressed in grams. Results from both hind paws were averaged. Assessment of mechanical hypersensitivity was performed between 10:00 and 17:00.

### Locomotion

Locomotor activity was measured using a Comprehensive Lab Animal Monitoring System (Columbus Instruments). Mice were acclimatized to the cages and individual housing for 24 h before a 12-h period (18:00–06:00) of automated recordings every 20 min. Movements in the x, y, and z axes were monitored during the third night after injection of antibodies, by recording the number of infrared beam breaks. Data are presented as total movement (total number of xy axis beam breaks) and rearing (number of beam breaks in the z axis) either over time or accumulated during the 12-h period. One or two control mice were included in each run, and the reference (control) group accumulated over the course of the locomotor experiments.

### Inverted grid test

Grip and muscular strength and forced movements were assessed by placing the mice on a surface (grid), which is then gradually turned upside-down. The turning takes 10 s, and then the latency to the mice losing their grip and falling off the grid is measured (with a cutoff of 10 s). The inverted grid test was performed 5 d after injection of saline or anti-CII mAbs. The investigators were blinded to the origin and treatment of the mice during the behavioral assessment.

### MMP activity

Mice injected with either saline or CII mAb cocktail received i.v. injection of MMPsense 680 (Galligan and Fish, 2012); 2 nmol in 150 µl PBS; PerkinElmer), an optically inert dye that becomes fluorescent in the presence of active MMPs, 24 h before sacrifice by decapitation. Paws were removed and scanned with Odyssey CLx (LI-COR), and signal intensity is presented normalized to saline-injected mice.

### Microarray expression analysis

Lumbar DRGs (L3–5) were dissected and total RNA extracted and purified using the RNeasy mini kit (Qiagen) according to the manufacturer's instructions. RNA concentration was measured using an ND-1000 spectrophotometer (NanoDrop Technologies), and RNA quality was evaluated using the Agilent 2100 Bioanalyzer system (Agilent Technologies). Total RNA (250 ng) from each sample was used to generate amplified and biotinylated sense-strand cDNA from the entire expressed genome according to the Ambion WT Expression Kit (P/N 4425209 Rev C 09/2009) and Affymetrix GeneChip WT Terminal Labeling and Hybridization User Manual (P/N 702808 Rev. 5; Affymetrix). GeneChip ST Arrays (GeneChip Mouse Gene 2.0 ST Array) were hybridized for 16 h in a 45°C incubator, rotated at 60 rpm. According to the GeneChip Expression Wash, Stain, and Scan Manual (PN 702731 Rev3; Affymetrix), the arrays were then washed and stained using the Fluidics Station 450 and finally scanned using the GeneChip Scanner 3000 7G. The raw data were normalized in the free software Expression Console provided by Affymetrix using the robust multiarray average (Li and Wong, 2001), followed by extraction of the expression levels of the genes of interest (*Fcgr1*, *Fcgr2b*, *Fcgr3*, and *Fcgr4*). Data are presented using a log2 scale with expression cutoff at 5.5.

Data have been deposited in the ArrayExpress database at EMBL-EBI under the accession no. E-MTAB-7853. In this article, we present data for the three control mice included in the study (individuals 254, 259, and 276).

### Quantitative real-time PCR

Mice were decapitated under volatile anesthesia, and ankle joints were collected fresh, trimmed from muscle and tendons, snap frozen, and stored at –70°C. For RNA extraction, joints were then pulverized using BioPulverizer (BioSpec) and briefly sonicated in TRIzol (Invitrogen) using an ultrasonic processor (EW-04714-51; Cole Parmer).

RNA was extracted according to the manufacturer's protocol and reverse transcribed to complementary DNA. For lumbar DRGs (L3–5), RNA extracted for microarray assay (described above) was used for reverse transcription. Quantitative real-time PCR (Applied Biosystems) was performed with hydrolysis probes, according to the manufacturer's instructions, to determine the relative mRNA levels. Predeveloped specific primer/probe sets for mouse chemokine Ccl2 (Mcp-1, Mm00441242_m1), inflammatory cytokines Tnf (Mm00443258_m1), Il1b (Mm00434228_m1), Il6 (Mm00446190_m1), mast cell proteases Mcpt4 (Mcp-4, Mm00487636-g1), Tpsb2 (Mcp-6, Mm01301240_g1), proinflammatory enzyme Cox2 (Mm00478374_m1), Mmp2 (Mm00439498_m1), Mmp9 (Mm00442991_m1), Mmp13 (Mm00439491-m1), and reference gene Hprt1 (Mm01545399_m1; all from Applied Biosystems) were used to determine threshold cycle values to calculate the number of cell equivalents in each sample with the standard curve method (Boyle et al., 2003). Data were normalized to Hprt1 values and expressed as relative expression units.

### smFISH

smFISH was performed as previously described with some modifications (Codeluppi et al., 2018). Mice were perfused with PBS under isoflurane anesthesia, and DRGs and sciatic nerves were collected and frozen in optimal cutting temperature (OCT) medium. After cryosectioning (10 µm), the sections were postfixed in 4% PFA (10 min at room temperature) and stored at –80°C until use. For hybridization, the sections were first permeabilized for 10 min with methanol in –20°C, cleared with 4% SDS, and after heat shock at 70°C for 10 min, incubated with 250 nM of fluorescent labeled probes for 4 h (Biosearchtech) at 37°C. After imaging, the tissues were counterstained with DAPI (Thermo Fisher Scientific) and NeuN-Alexa Fluor 488–conjugated antibody (ABN78A4; Millipore) for the specific labeling of neuronal cell bodies and TrkA antibody (AF1056; R&D Systems) to visualize axons. Sections were mounted with Prolong Gold (Thermo Fisher Scientific), and image stacks (0.3 µm) were acquired using a customized scanning microscope (Nikon TE). Images were analyzed using a custom Python script (pysmFISH python package). After background removal, a Laplacian-of-Gaussian was used to enhance the RNA dots that were defined as the local maxima above a threshold automatically calculated, after removal of connected components larger than dots. Quantification of single mRNA molecules as well as the cell diameter was done manually using ImageJ (National Institutes of Health), and data were plotted as number of RNA molecules according to cell size. The number of mRNA molecules per cell was then translated into a color gradient map and plotted according to cell size.

### Proteomic analysis

Lumbar DRGs (L3–5) collected from BALB/c mice were lysed by bead beating (Tissue Lyser; Qiagen) in triethylamonium bicarbonate buffer (Sigma-Aldrich) followed by protein thiol reduction. DRG lysates were digested to peptides using a filter-aided sample preparation method essentially as described (Wiśniewski et al., 2009) with minor modifications (Wiśniewski, 2016). The resulting peptides were recovered and proceeded to analysis (labeled and label-free). Peptide digests from one animal were labeled chemically using Tandem Mass Tag (TMT) 6-Plex reagents according to the manufacturer's instructions (Thermo Fisher Scientific), and the remaining peptide pool was analyzed unlabeled. To increase the proteome coverage, 100 µg of either TMT-labeled or unlabeled peptide samples were prefractionated using high pH reverse-phase liquid chromatography. The fractions were evaporated, reconstituted in 0.1% formic acid, and analyzed by high-resolution nanoLC-MS/MS on Q-Exactive Orbitrap mass spectrometers (Thermo Fisher Scientific) coupled to high-performance nanoLC systems (Dionex Ultimate-3000; Thermo Fisher Scientific) set up in a trap-and-elute configuration. For the TMT-labeled sample pool, data were also acquired using an identical nanoLC system interfaced to a TriBrid Orbitrap–based mass spectrometer (OrbiTrap Fusion; Thermo Fisher Scientific).

### DRG cell culture

DRGs (C1-L6) from WT and FcRγ chain$^{-/-}$ mice were extracted and placed in ice-cold Dulbecco's PBS until enzymatically dissociated with papain (1.7 mg/ml; 30 min at 37°C) followed by treatment with collagenase I (2 mg/ml) and dispase II (8 mg/ml; Sigma-Aldrich) enzyme mixture (30 min at 37°C). The cells were then gently triturated in Leibovitz's medium (L15) supplemented with 10% heat-inactivated FBS, 1% penicillin and streptomycin (Invitrogen), and 10 µM mitotic inhibitor (5-fluoro-2-deoxyuridine; Sigma-Aldrich). For CGRP release experiments, nerve growth factor (30 ng/ml; Sigma-Aldrich) was added into the medium. The cell suspension was plated on uncoated well plates for 1.5–2 h before being transferred to well plates precoated with poly-D-lysine and laminin (Sigma-Aldrich). The cells were maintained at 37°C in 5% $CO_2$ atmosphere, and the medium was replaced after 24 h, followed by changes every third day.

### CGRP release

After 6 d in culture, the medium was removed, and the cells were washed twice with Hepes buffer (25 mM Hepes, 135 mM NaCl, 3.5 mM KCl, 2.5 mM $CaCl_2$, 1 mM $MgCl_2$, 3.3 mM dextrose, and 0.1% [wt/vol] BSA, pH 7.4 with NaOH) and placed in new Hepes buffer for 30 min at 37°C (prestimulation). The Hepes buffer was collected for analysis of basal CGRP release. The cells were then incubated with CII-IC (0.1, 1, and 10 µg/ml), anti-CII mAb cocktail (1 µg/ml), CII antigen (1 µg/ml), or control IgG2b (1 µg/ml) in Hepes buffer or Hepes buffer alone for 30 min at 37°C (after stimulation), and the supernatant was collected for CGRP analysis. Capsaicin (50 nM; Sigma-Aldrich) in Hepes (10 min at 37°C) was used as a positive control. CGRP levels (pg/ml) in the supernatants were determined with an enzyme immune assay kit in accordance with the manufacturer's instructions (SPI-Bio; Bertin Pharma). The percent change before and after stimulation was calculated for each well.

### Calcium imaging

After 24 and 48 h in culture, the cells were loaded with Fluo-3 (4.4 µM; Life Technologies) for 30–40 min at room temperature (20–22°C). The cells were washed with modified Hepes buffer (145 mM NaCl, 3 mM KCl, 2 mM $CaCl_2$, 2 mM $MgCl_2$, 10 mM glucose, and 10 mM Hepes, pH 7.4 with NaOH) and then placed in the recording chamber and continuously perfused with bath solution (modified Hepes buffer) at a constant flow rate (1 ml/min). Calcium imaging was performed using a Nikon Diaphot inverted microscope with a diode laser (Cobolt dual calypso; Cobolt), 488-nm excitation, and a 40× oil-immersion objective. The change in emission (506 nm), i.e., intracellular calcium bound to Fluo-3, was recorded every 15 s using a photomultiplier tube (Bio-Rad MRC 1024). CII-IC (1 µg/ml) or control IgG2b (1 µg/ml) was applied for 3 min to the same cells in random order, with a minimum 10-min wash period between applications. At the end of each experiment, 50 mM KCl was applied for 1 min to detect functional neurons. All reagents were prepared from stock solutions and dissolved in modified Hepes buffer. Images were analyzed with ImageJ. In each image, capturing an average of 10 cells, all visible cells were chosen for analyses. Mean fluorescence intensity (F) for the region of interest, the cell bodies, was measured in each image. $F_0$ was calculated as the average mean intensity of the first five images in each series, and the data are presented as $F/F_0$. Cells were considered positive if the fluorescence signal increase was ≥20% compared with baseline and higher than three standard deviations.

### Electrophysiological recordings

Whole-cell voltage-clamp recordings were conducted in small DRG neurons (15–25 µm in diameter) at room temperature (20–22°C) within 24 and 48 h of culturing using a patch-clamp amplifier (Axo-Patch-200A; Molecular Devices). The recordings were filtered at 1 kHz, sampled at 4 kHz, and analyzed by using Clampex 10.4 software (Molecular Devices). Patch pipettes were pulled from borosilicate glass capillaries (Harvard Apparatus; 1.5-mm outer diameter, 0.86-mm inner diameter) using a vertical puller. The resistance of the patch pipettes was 4–5 MΩ when filled with internal solution (120 mM $K^+$-gluconate, 20 mM KCl, 1 mM $CaCl_2$, 2 mM $MgCl_2$, 11 mM EGTA, 10 mM Hepes, and 2 mM NaATP, pH 7.15 with Tris-base). Series resistance was not compensated for. DRG neurons were continuously perfused with bath solution (modified Hepes buffer, see Calcium

Table 1. **List of primary antibodies used for both IHC and immunocytochemistry (ICC) assays**

| Species/Tissue | Fixation | Primary antibody | Figure |
| --- | --- | --- | --- |
| Mouse: DRG (IHC) | Fresh; PFA 10 min RT | Rabbit anti-NeuN (1:100, Alexa Fluor 488–conjugated, ABN78A4; Millipore) | Fig. 4 D; Fig. 9, B–E; Fig. S4 A |
| Human: DRG (IHC) | PFA 2 h 4°C | | |
| Mouse: DRG (IHC), DRG cultures (ICC) | Fresh; acetone 10 min 4°C | Rat anti-FcγRI (2 µg/ml, gift from M.S. Cragg) | Fig. 5, A, B, E, and H; Fig. 6 A; Fig. S2, A, B, E, and F; Fig. S3 B |
| Mouse: Skin (IHC) | PFA-perfused; 4-h postfixation PFA | | |
| Mouse: DRG (IHC) | PFA-perfused; 24-h postfixation PFA | Rabbit anti-Iba1 (1:500, 019-19741; Wako) | Fig. 5 B |
| Mouse: DRG (IHC), DRG cultures (ICC) | Fresh; acetone 10 min 4°C | FcγRIIb (2 µg/ml, gift from M.S. Cragg) | Fig. 5, C, D, F, and I; Fig. 6 A |
| Mouse: Skin (IHC) | PFA-perfused; 4-h postfixation PFA | | |
| Mouse: DRG (IHC) | Fresh; acetone 10 min 4°C | Goat anti-TrkA (1:50, AF1056; R&D Systems) | Fig. 5, D, H, and I; Fig. S2 B |
| Mouse: Sciatic nerve (IHC) | Fresh; PFA 10 min RT | | |
| Mouse: Skin (IHC) | PFA-perfused; 4-h postfixation PFA | Rabbit anti-PGP (1:500, ab37188; Abcam) | Fig. 5, E and F |
| Mouse: DRG cultures (ICC) | Acetone 10 min 4°C | Mouse anti–β-III tubulin (1:1,000, G7121(A); Promega) | Fig. 6 A |
| Mouse: DRG (IHC) | PFA-perfused; 24-h postfixation PFA | Chicken anti-Vimentin (1:500, ab24525; Abcam) | Fig. S2 A |
| Mouse: DRG (IHC) | Fresh; acetone 10 min 4°C | Rat anti-FcγRIII (2 µg/ml, gift from M.S. Cragg) | Fig. S2 C |
| Mouse: DRG (IHC) | Fresh; acetone 10 min 4°C | Rat anti-FcγRIV (2 µg/ml, gift from M.S. Cragg) | Fig. S2 D |
| Mouse: DRG (IHC) | PFA-perfused; 24-h postfixation PFA | Goat anti-FcγRI (1:300, N-19, sc-7642; Santa Cruz Biotechnology) | Fig. S3 A |
| Mouse: DRG (IHC) | PFA-perfused; 24-h postfixation PFA | Rabbit anti-FcγRI (1:50, 50086-R001; Sino Biological) | Fig. S2 G; Fig. S3 C |
| Mouse: DRG (IHC) | PFA-perfused; 24-h postfixation PFA | Goat anti-FcγRI (1:100, AF2074; R&D Systems) | Fig. S2 H; Fig. S3 D |
| Human DRG (IHC) | PFA 2 h 4°C | Mouse anti-FcγRI (1:100, hCD64, clone 10.1, MCA756g; Serotec) | Fig. 9, B and C |
| Human DRG (IHC) | PFA 2 h 4°C | Mouse anti-FcγRIIIa/b (1:100, hCD16 PE/cy7-conjugated, clone 3G8; BioLegend) | Fig. 9, D and E |
| Human DRG (IHC) | PFA 2 h 4°C | Mouse anti-FcγRIIa (1:100, hCD32a, clone IV.3; Stem Cell Technologies) | Not shown |

RT, room temperature.

imaging protocol) at a constant flow rate (1–1.5 ml/min). CII-IC (1 µg/ml) and control IgG2b (1 µg/ml) were applied for 1 min, and capsaicin (0.5 µM) was applied at the end of each recording for 10 s as a control (4-min wash period between applications). Cells were accepted as having a resting membrane potential more negative than −40 mV and considered positive if the measured current was ≥20 pA and higher than three standard deviations. All reagents were prepared from stock solutions, dissolved in modified Hepes buffer, and applied via an 8-channel ValveLink 8.2 Controller application system (AutoMate Scientific).

### Western blot

For Western blot analysis, spleen and lumbar DRGs were harvested, snap frozen, and stored at −80°C until homogenized in lysis buffer. 5–10 µg of protein per well was loaded, separated by gel electrophoresis (4–12% Bis-Tris gel; Invitrogen), and transferred to nitrocellulose membranes. Nonspecific binding sites were blocked with 5% nonfat milk, and the membranes were probed with primary antibody overnight at 4°C (FcγRI, 0.1 µg/ml, 50086-R001; Sino Biological). After washing, the membranes were probed with secondary antibodies conjugated to HRP (Dako Antibodies), and the chemiluminescence signal (SuperSignal West Pico PLUS, 34580; Thermo Fisher Scientific) was detected by exposure to x-ray film (Fujifilm). Membranes were then stripped (Re-Blot plus; Millipore) and reprobed with primary antibody against β-actin (3700; Cell Signaling Technology) as a housekeeping protein reference. Quantification was performed using ImageJ.

## IHC and immunocytochemistry

For IHC, mice were deeply anesthetized and perfused with 4% PFA. Glabrous skin of the hind paws and lumbar DRGs were postfixed in PFA (4 and 24 h, respectively) and cryoprotected for 48 h in 30% sucrose in 0.01 M PBS at 4°C. For IHC with the FcγRIIb mAb (Table 1; Tutt et al., 2015), anesthetized animals were perfused with PBS before collection and snap freezing of tissues. Both fresh and perfused tissues were frozen in OCT and stored at −70°C until cryosectioning. Human DRGs (snap-frozen L4–5, collected from brain-dead subjects after asystole, $n = 4$, with the consent of first-tier family members) were collected at the University of Pittsburgh, shipped, and kept at −70°C until embedded in OCT. All procedures were approved by the University of Pittsburgh Committee for Oversight of Research and Clinical Training Involving Decedents and the Center for Organ Recovery and Education. Serial frozen sections of glabrous skin (20 µm) and DRGs (14 µm) were cut using a cryostat (NX-70; Thermo Fisher Scientific) and mounted on Superfrost plus glass slides. Fresh tissues were postfixed with acetone (10 min at 4°C) or PFA (10 min or 2 h at 4°C or room temperature; Table 1) immediately after sectioning. Primary cell cultures of DRG were also fixed in acetone (10 min at 4°C) 6 d after plating and stored in PBS at 4°C until analysis. Nonspecific binding was blocked using 5% normal serum in PBS (selection of serum dependent on species of secondary antibodies). Incubation with the primary antibodies (Table 1; Tutt et al., 2015) was performed overnight at room temperature. Anti-TrkA antibody was used for visualization of primary afferents because (1) a high percentage of joint nociceptors express TrkA (Mantyh et al., 2011) and (2) this antibody worked both on PFA- and acetone-fixed tissues. Immunoreactivity was visualized using Alexa Fluor–conjugated secondary antibodies (1:300; Invitrogen) or cyanine (Cy)-conjugated antibodies (1:300; The Jackson Laboratory). In human DRG sections, tyramide signal amplification (cy5-TSA kit, NE-L705A001KT; PerkinElmer) was performed using appropriate HRP-conjugated secondary antibodies (following the manufacturer's instructions). Prolong Gold antifade with DAPI (Life Technologies), was used for coverslipping, and images were collected using a confocal microscope (Zeiss LSM800) operated by LSM ZEN2012 (Zeiss) software. Figures were assembled in Illustrator CS6 (Adobe). To facilitate interpretation of the microscopic images, FcγR expression is always shown in green, with other markers depicted in red and DAPI staining in blue, independently of the secondary antibody used (images processed in ImageJ without modification of the capture settings).

## Statistical analysis

For comparing changes in behavior over time, repeated-measures two-way ANOVA was used followed by Bonferroni post hoc test. For differences in fluorescence, CGRP release, tactile thresholds, and locomotion with three groups or more, one-way ANOVA was used, followed by Bonferroni post hoc test. For differences in mRNA levels, tactile thresholds, and locomotion with two groups, Student's $t$ test was used. Arthritis and histological scores were compared using the Kruskal–Wallis test followed by Dunn's multiple comparison post hoc test, using GraphPad software. P values <0.05 were considered significant.

## Online supplemental material

Fig. S1 shows *Fcgr1* mRNA being expressed at the highest level compared with *Fcgr2b* and *Fcgr3* in smFISH experiments. The quantification of mRNA molecules is plotted per individual cell according to neuronal area and showed as intensity of color gradient. Fig. S2 shows that FcγRI protein does not colocalize with satellite cells (vimentin) or neuronal (TrkA) markers, but with antibodies from three different sources, it is present in resident macrophages in both BALB/c and C57BL/6 mouse DRG. Moreover, FcγRIII and FcγRIV proteins were not detected in mouse DRGs. Fig. S3 shows specificity control of the three antibodies used for detecting FcγRI protein in mouse DRG, since the resident macrophage staining disappears when using DRGs from FcRγ-chain$^{−/−}$ mice. One additional antibody used gave unspecific signal that was retained in DRGs from FcRγ-chain$^{−/−}$ mice. Fig. S4 shows that *Fcgr1–3* mRNA molecules were most highly expressed in fiber tracts of mouse DRGs in smFISH experiments.

## Acknowledgments

We thank Prof. Bo Rydqvist (Karolinska Institutet) for advice and assistance in electrophysiological recordings, Alice Wu (University of California, San Diego, CA) for the C5aR antagonist, Jeffrey Ravetch (The Rockefeller University, New York, NY) for FcγRIV$^{−/−}$ founder mice, the Strategic Research Programme in Diabetes at Karolinska Institutet for the services of the Metabolic Phenotyping Centre (Comprehensive Lab Animal Monitoring System recordings), and the University of Pittsburgh Health Science Core Research Facilities Genomics Research Core service, the Center of Organ Recovery and Education, and all the donors' families.

This work was supported by the Swedish Research Council (C.I. Svensson, R. Holmdahl, K.S. Nandakumar, M. Collin, J.T. Lanner, and F. Wermeling), Swedish Foundation for Strategic Research (C.I. Svensson and R. Holmdahl), Knut och Alice Wallenberg Foundation (C.I. Svensson and R. Holmdahl), Ragnar Söderberg Foundation (C.I. Svensson), Torsten Söderberg Foundation (M. Collin), Åke Wiberg Foundation (M. Collin), Alfred Österlund Foundation (M. Collin), Gyllenstierna-Krapperup Foundation (M. Collin), Konung Gustaf V's 80-year foundation (D.B. Bas, K.S. Nandakumar, M. Collin, and F. Wermeling), Swedish Arthritis Association (K.S. Nandakumar, J.T. Lanner, and F. Wermeling), Hansa Medical AB (M. Collin), the Royal Physiographic Society in Lund (M. Collin), funds from Karolinska Institutet (K.S. Nandakumar, J. Sinclair, A. Bersellini Farinotti, J.T. Lanner, and F. Wermeling), the Canadian Institutes of Health Research (G237818/CERC09/CIHR; L. Diatchenko), and a grant from the Government of Guangdong Province (201001Y04675344; R. Holmdahl). The funders had no role in the preparation of the manuscript or in the decision to publish.

Hansa Medical AB holds patents for using EndoS as a treatment for antibody-mediated diseases. M. Collin is listed as one of the inventors on these patents and has a royalty agreement with Hansa Medical AB. M. Collin also holds patents for the biotechnological use of EndoS. The other authors declare no competing financial interests.

Author contributions: Conceptualization, A. Bersellini Farinotti, G. Wigerblad, D. Nascimento, D.B. Bas, K.S. Nandakumar, L. Diatchenko, S. Codeluppi, R. Holmdahl, and C.I. Svensson; Methodology, A. Bersellini Farinotti, G. Wigerblad, D. Nascimento, D.B. Bas, K.S. Nandakumar, K. Sandor, B. Xu, L.E. Borm, F. Wermeling, B. Heyman, M. Collin, K. Kultima, B. Heyman, J.M. Jimenez-Andrade, S. Codeluppi; Investigation, A. Bersellini Farinotti, G. Wigerblad, D. Nascimento, D.B. Bas, C. Morado Urbina, K. Sandor, S. Abdelmoaty, M.A. Hunt, K. Ängeby Möller, A. Baharpoor, L. Zhang, J. Sinclair, L.E. Borm, K. Kultima, J.M. Jimenez-Andrade, S. Codeluppi; Resources, K.S. Nandakumar, B. Xu, I. Khmaladze, L. Zhang, M.S. Cragg, J. Lengqvist, A.-J. Chabot-Doré, L. Diatchenko, Inna Belfer, K. Kultima, M. Collin, F. Wermeling, B. Heyman, R. Holmdahl, and C.I. Svensson; Writing – original draft, A. Bersellini Farinotti, G. Wigerblad, D. Nascimento, D.B. Bas, K.S. Nandakumar, M.A. Hunt, R. Holmdahl, and C.I. Svensson; Writing – review and editing, all authors; supervision, K.S. Nandakumar, K. Jardemark, L. Diatchenko, J.T. Lanner, K. Kultima, S. Codeluppi, R. Holmdahl, and C.I. Svensson; Funding acquisition, R. Holmdahl and C.I. Svensson.

Submitted: 30 August 2018
Revised: 20 January 2019
Accepted: 24 April 2019

## Supplemental material

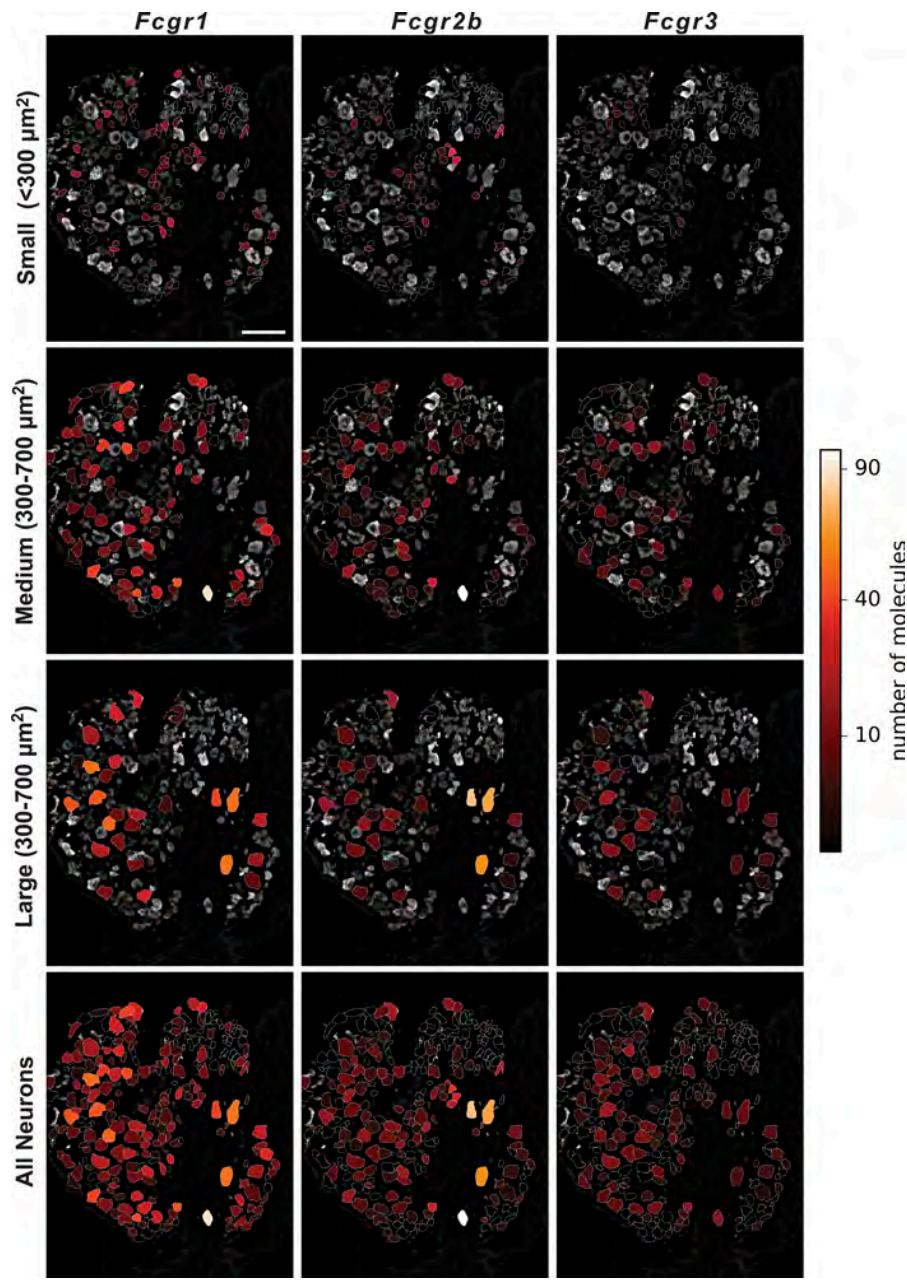

Figure S1. **Quantification from smFISH of mRNA molecules of *Fcgr1*, *Fcgr2b*, and *Fcgr3* in mouse DRG.** Related to Fig. 4. Scale bar represents 100 µm. *Fcgr1–3* molecules are plotted per individual cell according to neuronal area and showed as intensity of color gradient. *Fcgr1* is shown expressed at the highest level.

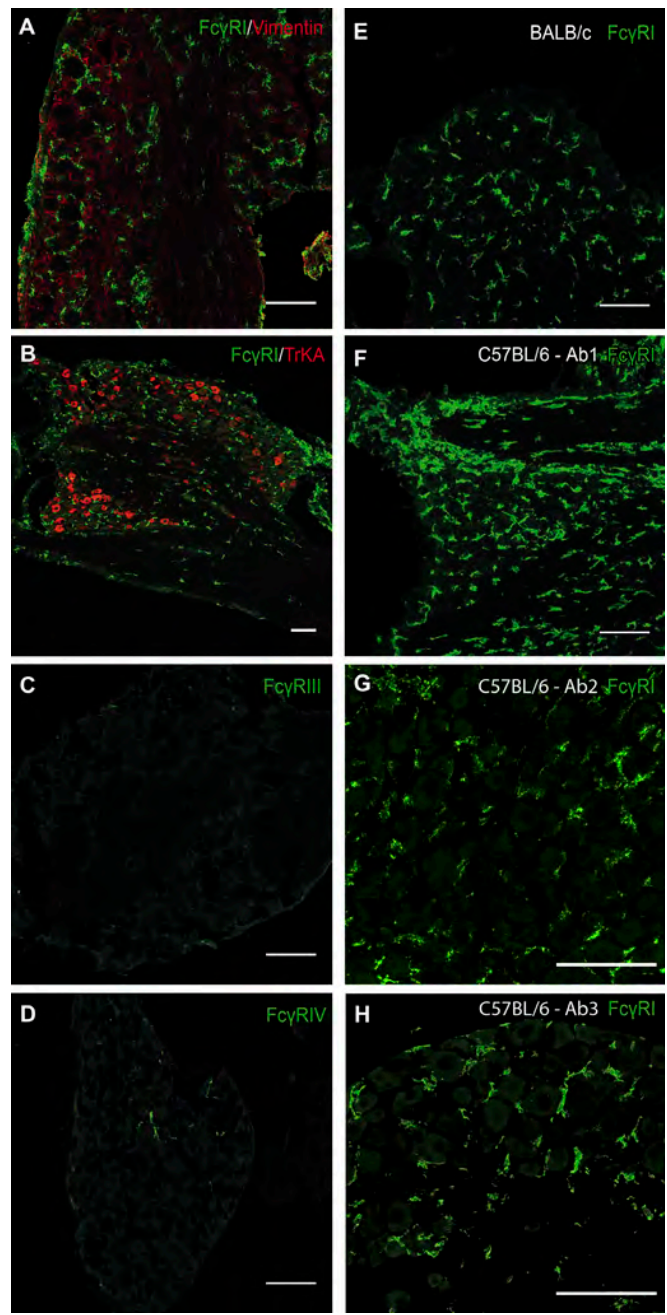

Figure S2. **FcγRI expression in BALB/c and C57BL/6 mice DRGs and lack of expression of FcγRIII and FcγRIV in mouse DRG.** Related to Fig. 5. **(A and B)** IHC shows lack of colocalization between FcγRI staining and vimentin and TrkA. **(C and D)** No immunoreactivity for FcγRIII and FcγRIV was detected in BALB/c mouse DRG. **(E)** FcγRI immunoreactivity is present in resident macrophages of BALB/c mouse DRG. **(F–H)** Antibodies from three different sources (M.S. Cragg, Sino Biological, and R&D Systems) showed FcγRI immunoreactivity in resident macrophages in DRGs from C57BL/6 mice. Scale bars represent 100 µm.

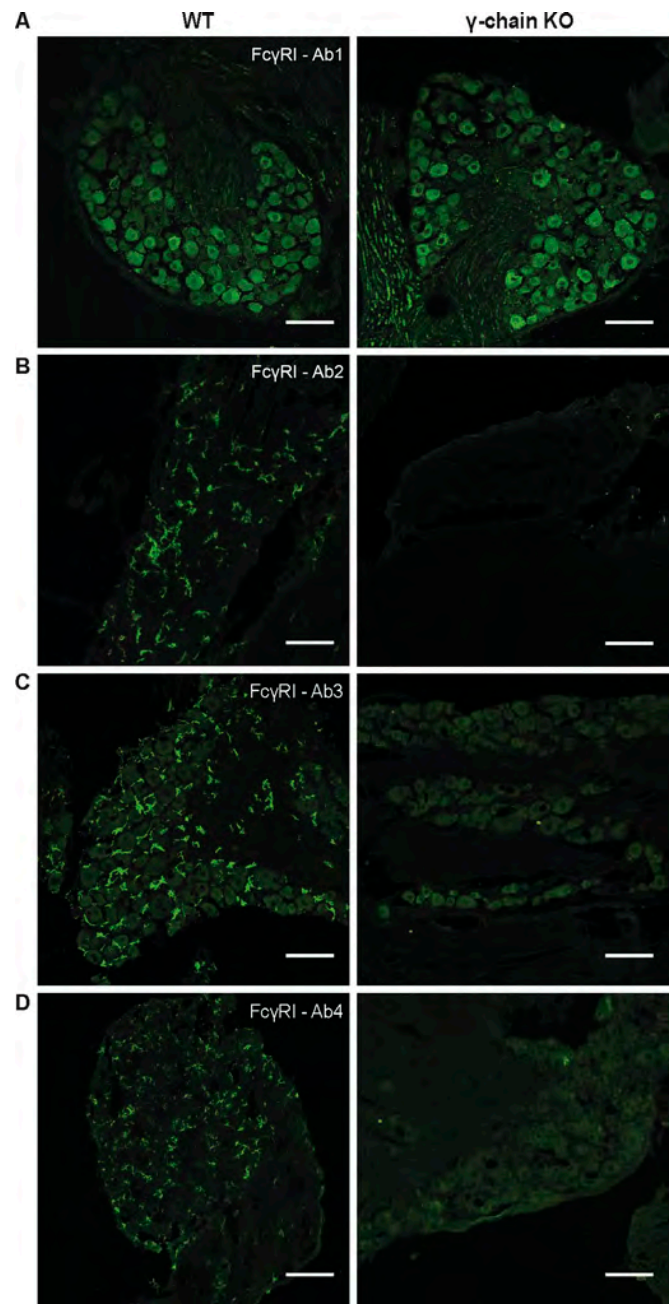

Figure S3. **Specificity control of several anti-FcγRI antibodies used for IHC in mouse DRG.** Related to Fig. 5. **(A)** FcγRI immunoreactivity is present in neuronal cell bodies of BALB/c mouse DRG when using an antibody from Santa Cruz Biotechnology, but the signal is retained in DRGs from FcRγ-chain$^{-/-}$ mice, indicating nonspecific binding. **(B–D)** Antibodies from three different sources (M.S. Cragg, Sino Biological, and R&D Systems) showed FcγRI immunoreactivity in resident macrophages in BALB/c mouse DRGs, and the signal is absent in DRGs from FcRγ-chain$^{-/-}$ mice, indicating specific binding. Scale bars represent 100 µm.

**Bersellini Farinotti et al.**
Immune complexes induce pain via neuronal FcγRI

**Journal of Experimental Medicine**  S3
https://doi.org/10.1084/jem.20181657

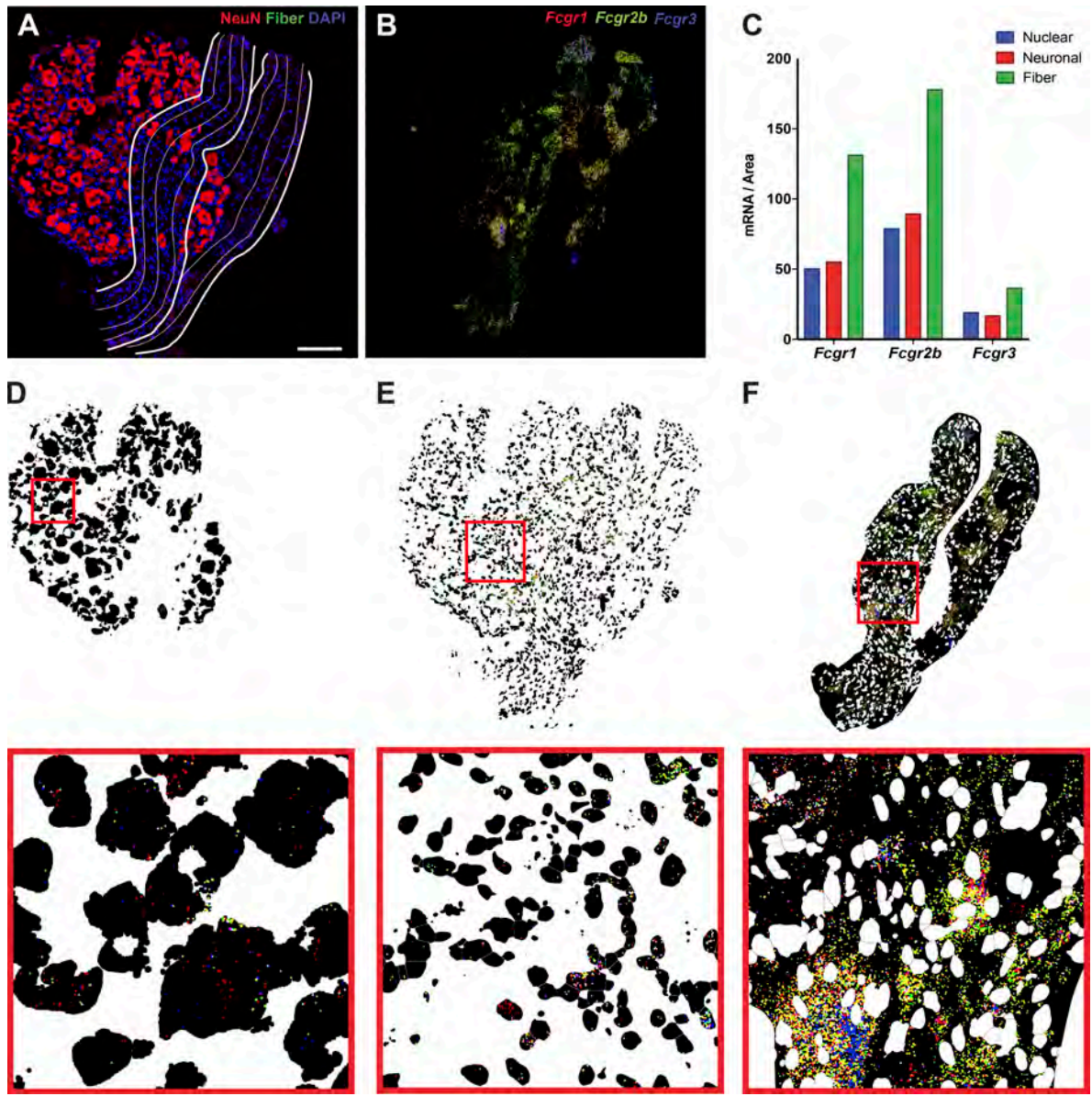

Figure S4. **smFISH quantification of *Fcgr1*, *Fcgr2b*, and *Fcgr3* mRNA molecules in neuronal, nuclear, and fiber tract areas of mouse DRGs.** Related to Figs. 4 and 5. **(A)** Representation of neuronal (NeuN), nuclear (DAPI), and fiber tract (nonneuronal and nonnuclear) areas in mouse DRG. **(B)** smFISH detection of *Fcgr1–3* in mouse DRGs. **(C–F)** Quantification of *Fcgr1–3* in neuronal, nuclear, and fiber tract areas. *Fcgr1–3* are most expressed in the fiber tracts of mouse DRG. Panels D, E, and F are masked versions of B, representing the neuronal, the nuclear, or the fiber regions of the whole DRG, respectively. Scale bar represents 100 µm.